\documentclass[twocolumn]{aa}
\usepackage{graphicx}
\usepackage{natbib}

\usepackage{txfonts}
\begin{document}
   \title{X-rays from jet-driving protostars and T Tauri stars}

   \author{Manuel G\"udel
          \inst{1}
          \and
	  Alessandra Telleschi
	  \inst{1}
	  \and 
	  Marc Audard
	  \inst{2}
	  \and
	  Stephen L. Skinner
	  \inst{3}
	  \and
	  Kevin R. Briggs
	  \inst{1}
	  \and
	  Francesco Palla
	  \inst{4}
	  \and
	  Catherine Dougados
	  \inst{5}
          }

   \offprints{Manuel G\"udel}

   \institute{Paul Scherrer Institut, W\"urenlingen and Villigen,
              CH-5232 Villigen PSI, Switzerland\\
              \email{guedel@astro.phys.ethz.ch}
	 \and
	      Columbia Astrophysics Laboratory,
              Mail Code 5247, 550 West 120th Street,
              New York, NY 10027,
              USA     
	 \and 
	     CASA, 389,
             University of Colorado,
             Boulder, CO 80309-0389,
             USA   
	 \and
	     Osservatorio Astrofisico di Arcetri,
             Largo Enrico Fermi, 5,
             I-50125 Firenze,
             Italy     
 	 \and
             Laboratoire d'Astrophysique de Grenoble,
             Universit\'e Joseph Fourier - CNRS,
             BP 53,
             F-38041 Grenoble Cedex,
             France
            }

   \date{Received 1 June 2006; accepted 25 August 2006}

  \abstract
   {} 
   {We study  jet-driving protostars and T Tau stars to characterize their X-ray emission. We seek
   soft spectral components that may be due to shock emission,
   and search for soft, shock-induced emission displaced from the stellar position. }
   {We study two stellar samples, the first consisting of lightly absorbed T Tau stars with strong jets, 
   the other  containing protostars with disks seen nearly edge-on. The former sample was observed in the {{\it XMM-Newton Extended Survey of
    the Taurus Molecular Cloud}} (XEST), while {{\it Chandra}} archival data provided observations of the latter.}
   {We confirm the previously identified peculiar spectrum of DG Tau A and find similar phenomenology in GV Tau and DP Tau, suggesting
   a new class of X-ray  spectra.   These  consist of a lightly absorbed, very soft 
   component and a strongly absorbed, very hard component. The latter is flaring while little variability 
   is detected in the former. The absorption of the harder component is about an order  of magnitude higher 
   than expected from the optical extinction assuming a standard gas-to-dust mass ratio.  For the
   absorbed protostars, only the hard, stellar X-ray component is found.
   }
   {The flaring hard component represents active coronal emission. 
   Its strong absorption is attributed to mass inflow from the accretion disk. The optical extinction is small
   because the dust has sublimated at larger distances. The weakly absorbed soft component cannot originate from the 
   same location. Because the stars drive strong jets, we propose that the X-rays are generated in shocks in the jets. 
   We find that for the three peculiar X-ray sources, the luminosity of the 
   soft component  roughly scales with the equivalent width of the [O~I] $\lambda$~6300 line formed in the jets, and with 
   the mass outflow rate. In the more strongly obscured protostars, the soft component is
   entirely absorbed, and only the hard, coronal component penetrates the envelope or the near-edge-on disk.}

   \maketitle
%

\section{Introduction}\label{introduction}

Pre-main sequence stars show various signs of accretion and 
outflow,  such as circumstellar disks detected at
radio and millimeter wavelengths (e.g., \citealt{simon00}), 
molecular outflows observed in molecular lines (e.g., \citealt{bachiller96} for a review),
and accompanying optical (e.g., \citealt{hirth97, eisloeffel98}) and radio jets
(e.g., \citealt{anglada95}).  The most visible manifestation of outflows are 
the optically visible jets that  may be excited close to the star,
in internal shocks along the mass stream, or at distances up to several 
parsecs as the fast stream encounters the interstellar medium where
it shock-ionizes the gas in Herbig-Haro objects (for a review of Herbig-Haro 
flows, see \citealt{reipurth01}). Under ideal circumstances (low extinction, 
strong ionization), jets can be identified at distances as close as 0.1\arcsec\ 
to the star \citep{bacciotti02}. The same compact jets are also routinely detected
at radio wavelengths, where the emission mechanism  is thought 
to be bremsstrahlung from the shock-heated gas \citep{rodriguez95, anglada95}.
Radio brightness temperatures suggest overall gas temperatures of order  $10^4$~K. 
This picture is ambiguous, however, as a number of non-thermal jets have been
suggested from radio polarization or synchrotron-like spectral shapes 
\citep{yusef90, curiel93, hughes97, ray97, smith03, loinard05}. 
Magnetic fields may thus play a role not only in the launching of jets, but in 
their propagation as well.

Two principal families of models for the jet-launching mechanism
have been proposed:
The magnetocentrifugal mechanism \citep{blandford82, koenigl00} posits
that outflows are launched on the disk along outward-bent, open magnetic 
fields owing to the increasing centrifugal forces along 
the field lines that are co-rotating with their disk footpoints.
In contrast, the X-wind model \citep{shu00} assumes that outflows are 
launched from magnetized regions between the star and the disk. Magnetohydrodynamic
simulations show that magnetic reconnection and subsequent plasmoid ejection
may lead to field-aligned outflows \citep{hayashi96}. 

Accretion and outflow processes are prone to producing X-rays,
given that shocks with shock jump velocities of order several  hundred km~s$^{-1}$ 
are possible and  sometimes observed. Accretion onto the star 
essentially extracts gravitational energy and transforms it into thermal energy 
through shocks. The relevant theory has been presented by
\citet{ulrich76}, \citet{calvet98} and \citet{lamzin99}. Material falling along
magnetic fields toward the stellar surface with nearly free-fall velocity, $v_{\rm ff}$,
develops strong shocks near or within the photosphere, heating plasma to
temperatures of $T \approx 3v_{\rm ff}^2\mu m_p/16k \approx 
5.2\times 10^6 M/R~$~[K] where $M$ and $R$ are the stellar mass and the stellar
radius, respectively, in solar units, $\mu$ is the mean molecular weight,
$m_p$ is the proton mass, and $k$ is the Boltzmann constant. For T Tau stars, heating to a few MK
is thus entirely plausible (as, e.g., proposed for TW Hya, \citealt{kastner02}), 
but  the temperatures in excess of 10~MK   that often dominate X-ray emitting plasmas 
in T Tau stars  \citep{skinner03, preibisch05} cannot be due to accretion shocks. Rather, 
high electron temperatures in pre-main sequence stars are conventionally 
attributed to magnetized plasma trapped in  corona-like or large-scale ``magnetospheric'' 
magnetic fields around the stars. 

Shocks also form in jets and outflows, in particular in Herbig-Haro (HH)
objects. The relevant theory and a simple model have been discussed  by \citet{raga02}. 
The strong-shock temperature can be expressed as $T\approx 1.5\times 10^5v_{\rm 100}^2$~K 
(for fully ionized gas) where $v_{100}$ is the shock  speed relative to a target  
in units of 100~km~s$^{-1}$. Jet speeds are typically of order $v = 300-500$~km~s$^{-1}$ 
\citep{eisloeffel98, anglada95, bally03},
allowing for shock speeds of similar magnitude. If a flow 
shocks a standing medium at 400~km~s$^{-1}$, then $T \approx 2.4$~MK. 

Observations have explicitly demonstrated that faint, soft X-ray emission originates from
some  HH objects \citep{pravdo01, pravdo04, favata02, tsujimoto04, pravdo05, grosso06}. \citet{bally03}
 used a {\it Chandra} observation to show that  X-rays form within
an arcsecond of the protostar L1551 IRS-5 while the star itself
is too  heavily obscured to be detected. As this example illustrates, the jet-launching region 
of powerful protostellar jets is usually inaccessible to optical, near-infrared or 
X-ray studies of protostellar outflow sources due to excessive absorption. However, there are a number
of optically-revealed, classical T Tau stars that drive appreciable
jets and outflows \citep{hirth97}. Their jets  were  initially  dubbed 
``micro-jets'' although recent studies have shown them to reach parsec scales 
and the mass loss rates  compete with those of more embedded sources 
\citep{mcgroarty04}.

One of the most prominent examples, the classical T Tau star DG Tau A, has recently 
obtained scrutiny with high-resolution {\it Chandra} X-ray observations. \citet{guedel05}
discovered a new type of X-ray spectrum in which a lightly
absorbed spectral component from a rather cool (2--3~MK) plasma is complemented
by a strongly absorbed component from a very hot (20--30~MK)  plasma. The latter
was interpreted as being related to the usual coronal/magnetospheric source often
seen in T Tau stars, subject to some excess absorption. In the light of this 
interpretation, the cooler plasma is unlikely to be located close to the
magnetospheric component because the source of absorption would 
likely affect it as well.  The strong jets and high mass-loss rates of DG Tau A  
motivated the suggestion that the soft X-rays are produced
in shocks in the acceleration region of the jet (the ``base'' of 
the jet), which is unresolved in {\it Chandra} or {\it XMM-Newton} 
images. Complementary  evidence was seen in faint and very soft X-ray emission along 
the optical jet, outside the stellar point-spread function and co-spatial with a strong
bow-shock feature seen in the optical. The spectral distribution of the counts in the
jet source was reminiscent of the count distribution in the soft stellar component.
A spectrum requiring two absorbers was also reported for a proplyd in the  
the Orion Nebula Cluster, with a variable hard component but a constant soft component \citep{kastner05}.
\citet{skinner06} have detected a similar two-component spectrum in the strongly accreting FU Ori, the
prototype of FU Ori-type stars. The cooler, weakly absorbed component, however,
reached temperatures more common to T Tau stars ($\approx 8$~MK) and was therefore
identified with a coronal component, while the very hot component was again 
strongly absorbed.

The recently conducted {\it XMM-Newton Extended Survey of the Taurus Molecular Cloud} (XEST henceforth; 
\citealt{guedel06})  offers unique access to numerous 
stellar objects in the Taurus Molecular Cloud (TMC). It covers  most stellar
concentrations in the TMC molecular filaments and offers high-quality spectra and light curves
for any ``typical'' T Tau star in the field of view. XEST also covers a number of 
well-studied jet-driving T Tau stars and protostars. The observations confirm the initial
discovery of the anomalous DG Tau A and add substantial evidence for a
jet-related origin. We find further clear examples of these 
{\it two-absorber X-ray} sources (TAX sources henceforth).  
The present paper describes these new observations and presents 
a detailed discussion on the possible origin of the X-ray emission. For a comparison,
we also present X-ray spectra of jet-driving sources that are deeply absorbed
by their thick circumstellar disks seen nearly edge-on. The soft component,
if present, would thus be absorbed {\it unless} it is displaced sufficiently from the
star to make it separate from the stellar X-ray image. These complementary data are based on
high-resolution {\it Chandra} observations.

\begin{table*}[t!]
\centering
\caption{Principal targets and observations}
\label{obslog}
\begin{tabular}{lllll}
\hline
\noalign{\smallskip}
            Parameter                  &        DG Tau A                &    GV Tau                        & DP Tau                           &    CW Tau        \\
            \noalign{\smallskip}
            \hline
            \noalign{\smallskip}
	    XEST number                &  02-022                        &	13-004                     &  10-045                          &   20-046 \\	
	    ObsID  {\it (XMM-Newton})  &  0203540201                    &	0203541301                 &  0203542201                      &  0203542001  \\	
	    Start time  (UT)           &  2004-08-17\ 06:08:10          &  2004-08-25\ 11:11:23            &  2005-03-05\ 05:56:38            &  2004-09-12\ 07:04:43      \\ 		
	    End time (UT)              &  2004-08-17\ 17:32:46          &  2004-08-25\ 19:59:19            &  2005-03-05\ 14:44:30            &  2004-09-12\ 15:52:37    \\
	    Exposure time  (s)         &  41076		                &  31676                           &  31672                           &  31674  \\
	    Boresight RA~(J2000.0)     &  04h 27m 19.6s                 &  04h 29m 52.0s                   &  04h 42m 20.9s                   &  04 14 12.9     \\  
	   Boresight $\delta$~(J2000.0)&  $26\deg\ 09\arcmin\ 25\arcsec$&  $24\deg\ 36\arcmin\ 47\arcsec$  &  $25\deg\ 20\arcmin\ 35\arcsec$  &  28 12 12  \\ 
	    Star: RA~(J2000.0)$^a$     &  04h 27m 04.70s                &  04h 29m 23.73s	           &  04h 42m 37.70s                  &  04h 14m 17.00s  \\
	   Star: $\delta$~(J2000.0)$^a$&  $26\deg\ 06\arcmin\ 16\farcs 3$&  $24\deg\ 33\arcmin\ 00\farcs 3$ &  $25\deg\ 15\arcmin\ 37\farcs 5$& $28\deg\ 10\arcmin\ 57\farcs 8$   \\  
            \hline
\end{tabular}
\vskip 0.5truecm
\begin{tabular}{lllll}
\hline
\noalign{\smallskip}
            Parameter                  &   DG Tau B                     &    FS Tau B                      &  HH 34 IRS                       &    HH 111 IRS              \\
            \noalign{\smallskip}
            \hline
            \noalign{\smallskip}
	    XEST number                &  C3-2                          &	11-054 = C2-1              &  ...                             &   ...        \\	
	    ObsID {\it (Chandra})      &  4487                          &  4488                            &  4489                            & 4490         \\
	    Start time  (UT)           &  2004-01-11\ 02:58:51          &  2003-11-08\ 12:57:58            &  2003-12-05\ 08:59:23            & 2004-11-21\ 17:05:50    \\
	    End time (UT)              &  2004-01-11\ 11:52:21          &  2003-11-08\ 21:56:34            &  2003-12-05\ 17:38:31             & 2004-11-22\ 01:36:42     \\
   	    Exposure time  (s)         &   29717                        &  29674                           &  28590                           & 28795       \\
	    Boresight RA~(J2000.0)     &   04h 27m 02.3s                &  04h 22m 00.2s                   &  05h 35m 32.07s                  & 05h 51m 54.97s       \\
	   Boresight $\delta$~(J2000.0)& $26\deg\ 04\arcmin\ 56\arcsec$ &  $26\deg\ 58\arcmin\ 07\arcsec$  & $-06\deg\ 26\arcmin\ 50\farcs 0$ & $02\deg\ 48\arcmin\ 55\farcs 3$ \\
	   Star: RA~(J2000.0)$^b$      &  04h 27m 02.55s	        &  04h 22m 00.70s                  & 05h 35m 29.84s                   & 05h 51m 46.31s              \\
	   Star: $\delta$~(J2000.0)$^b$& $26\deg\ 05\arcmin\ 30\farcs 9$&  $26\deg\ 57\arcmin 32\farcs 50$ & $-06\deg\ 26\arcmin\ 58\farcs 40$& $02\deg\ 48\arcmin 29\farcs 72$ \\
         \hline
\end{tabular}
\begin{list}{}{}
\item[$^{\mathrm{a}}$] For origin of coordinates, see \citet{guedel06}  
\item[$^{\mathrm{b}}$] For origin of coordinates, see  Tables~\ref{proto} and \ref{GVcoords} 
\end{list}
\end{table*}

The plan of the paper is as follows. In Sect.~\ref{sect:targets}, we briefly describe our 
data analysis and  introduce the targets of this study. Sec.~\ref{sect:results} presents the
results that are further discussed in Sect.~\ref{sect:discussion}. We conclude in
Sect.~\ref{sect:summary}.

\section{Targets and observations}\label{sect:targets}

\subsection{Data reduction}

The observations reported here were obtained as part of the XEST project and
have been subject to a data reduction procedure that is described in 
detail in \citet{guedel06}. In brief, the {\it XMM-Newton} \citep{jansen01} 
data have been obtained using all three EPIC cameras (MOS and PN; \citealt{turner01}, 
\citealt{strueder01}), with the medium
filter inserted. We applied extraction regions that maximize the signal-to-noise ratio;
their radii amount to 22\arcsec, 15\arcsec, 13\arcsec, and 12\arcsec\ for our targets 
DG Tau A, GV Tau, DP Tau, and CW Tau, respectively. Standard response matrices
distributed by the {\it XMM-Newton} project were used,  and ancillary response files
were created for each source. An observing log
is given in Table~\ref{obslog}.  For DG Tau A and GV Tau, we also extracted the simultaneous 
light curves from the Optical Monitor (OM; \citealt{mason01}). The OM data reduction is described in \citet{audard06}.

For spectral analysis, we used the XSPEC software package \citep{arnaud96}. We applied
the {\sl vapec} thermal collisional ionization equilibrium models that include
emission lines and bremsstrahlung continua, combined with the photoelectric
absorption model, {\sl wabs}, based on absorption cross sections by \citet{morrison83}.

The {\it Chandra} exposures described here were reduced using standard strategies for ACIS.
The observations of DG Tau AB, FS Tau AB (= Haro 6-5 AB), HH~34, and HH~111 used the 
``Very Faint Mode'' of ACIS to minimize background contributions, while the GV Tau (= Haro 6-10 AB)
observation used the ``Faint Mode''. The data were reduced in CIAO vers. 3.0.2 following the 
standard analysis threads\footnote{http://cxc.harvard.edu/ciao/guides/}. These procedures 
included corrections for charge transfer inefficiency and afterglow, and selection
of good time intervals.  Spectra were extracted from level 2 event
files with  {\sl specextract} which also produces response  matrices and ancillary response files.

\subsection{Jet-driving T Tau stars}

The XEST survey contains several jet-driving, optically revealed classical T Tau stars 
\citep{hirth97},  in particular \object{DG Tau A}, \object{GV Tau AB} = \object{Haro 6-10 AB}, 
\object{DP Tau}, \object{CW Tau},
\object{XZ Tau}, \object{UZ Tau}, and \object{DD Tau}. One further object, \object{HN Tau}, was observed with the {\it Chandra} 
HRC-I but was recorded only as a very faint source \citep{guedel06}.

At least three XEST targets, namely DG Tau A, GV Tau A, DP Tau, and possibly a fourth, CW Tau, reveal
the same anomalous spectral phenomenology previously described for the first object  
\citep{guedel05}. Further, UZ Tau reveals a peculiarly soft
spectrum, while DD Tau shows a very hot source that is subject to an anomalously high
photoelectric absorption, judged from the visual extinction which is relatively small
\citep{guedel06}. The present paper concentrates on a discussion of the first three
unambiguous cases.

\begin{table*}[t!]
\centering
\caption{Fundamental parameters of optically revealed stars}
\begin{tabular}{lrrrrrrrrrr}
\hline
{\rm Parameter}                               &DG Tau A& Ref$^a$&GV Tau A&Ref$^a$& GV Tau B&Ref$^a$&DP Tau&Ref$^a$&CW Tau  &Ref$^a$\\
\noalign{\smallskip}
\hline
\noalign{\smallskip}
Spectral type                       	          &K5-M0&4, 5, 7, 8, 15&K3-7& 4, 8 & ...    &... &M0-0.5 & 5, 8     & K3       &  7     \\
YSO class                           	          & II	& 4, 8	  & I	&4, 6, 8&   I	    &... & II	 &4, 6, 8, 18&	   II  &  4, 6, 8  \\
                                    	          &FS$^b$& 6, 18  &	&	&	    &	 &	 &          &	       &     \\
T Tau type                           	          &CTTS &5, 17	  &CTTS & 8	& ...	    &... & CTTS  & 5	    & CTTS     & 17 \\
$T_{\rm eff}$ (K)                     	          &4205 & 7	  &4730 & 7	& 5150      & 12 & 3778  & 14	    & 4730     &  7 \\
                                    	          &4775 & 8	  &4000 & 8, 12 &	    &	 & 3800  & 8	    &	       &    \\
Mass ($M_{\odot}$)                    	          &0.9-1.8&7, 8$^c$&0.7-2.1&4, 8$^c$&  2    & 12 &0.52-0.53& 4, 8$^c$& 1.4      & 4$^c$\\
                                    	          &0.67 &  15	  &	&	&	    &	 &	 &	    &	       &    \\
                                    	          &0.88 &  5	  &	&	&	    &	 &	 &	    &	       &      \\
Radius  ($R_{\odot}$)                 	          &2.5$^d$& ...   &2.8$^d$& ... &  3.1      & 12 &1.05$^d$&   ...   & 1.6$^d$  & ...  \\
W([O~I]) (\AA)  		   	          &11-22& 15      & 3.8 & 2	& ...       &... & 0.7   & 2 	    &  3.5     &  2  \\
         			    	          &     &         &     &  	&           &    &       &    	    &  7.9     &  15  \\
$L_*\ (L_{\odot})$                  	          & 3.62& 8	  & 1.8 & 8	& ...	    &... & 0.2   & 14	    &  1.1     &  7 \\                                    	         
                                                  & 1.7 & 7, 15   & 1.3 & 12    &           &    & 0.3   & 5        &  0.76    & 5  \\
						  &1.15 & 5       &     &       &           &    &       &          &          &   \\       
$L_{\rm bol}\ (L_{\odot})$          	          & 6.4 & 4	  & 6.98& 4	& 6.1	    & 12 & 0.7   & 4, 8     &  2.7     &  4 \\
                                    	          & 8	& 1	  &	&	&	    &	 &	 &	    &	       &    \\
$A_{\rm V}$ (mag)                   	          & 1.41& 5	  & 3.3 & 11	& 49	    & 12 & 1.26  & 5	    & 3.4      & 15   \\
                                    	          & 2.20& 10	  & 5.6 & 12	&	    &	 & 1.46  & 4	    &  2.29    &  4  \\
                                    	          & 3.32& 8	  & 12.1& 8	& 35	    & 13 & 6.31  & 8	    &	       &  \\
                                    	          & 3.2 & 15	  &	&	&	    &	 &	 &	    &	       &    \\
$A_{\rm J}$ (mag)                   	          & 0.36& 7	  & ... & ...	& ...	    & ...& 0.41  & 14	    & 0.55     & 7 \\
$\log\dot{M}_{\rm acc}\ (M_{\odot}{\rm yr}^{-1})$ &-7.34& 5       &-6.71& 8     & ...       & ...& -8.50 & 5        & -7.99    & 5  \\
                                                  &-6.13& 8       &     &       &           &    & -6.92 & 8        &          &   \\
                                                  &-5.7 & 15      &     &       &           &    &       &          &          &   \\
$\log\dot{M}_{\rm out}\ (M_{\odot}{\rm yr}^{-1})$ &-6.19& 8       &-6.54& 8     & ...       &... & -7.42 & 8        & -7.1     & 15   \\
                                                  & -6.5& 15      &     &       &           &    &       &          &          &   \\
                                                  & -6.1& 10      &     &       &           &    &       &          &          &    \\
                                                  & -6.6& 16      &     &       &           &    &       &          &          &    \\
$M_{\rm disk}\ (M_{\odot})$                       & 0.02& 6       &0.003& 6     & ...       &... &$<$0.0005&  6     &  0.002   &  6 \\
age (Myr)                                         &1.3-2.2&7, 8$^c$&0.89-0.98&4, 8$^c$& ... &... &1.5-7.2&8, 14$^c$ & 7        & 4$^c$  \\
$v_{\rm rad}$ (km~s$^{-1}$)                       & 250 & 2       &80-300& 2    & ...       &... & 90-110& 2        & 110      & 2   \\
$v_{\rm pm}$ (km~s$^{-1}$)                        & 194 & 3       &...  &  ...  & ...       &... & ...   & ...      & ...      &  ...   \\
$\Delta v_{\rm rad}$ (km~s$^{-1}$)                & ... &  ...    &   90& 2     & ...       &... & 86    & 2        & ...      & ...   \\
$P_{\rm rot}$ (d)                                 & 6.3 & 19      & $<5.6^e$&   & ...       &... &$<2.8^e$ & ...    & $<2.9^e$ & ...   \\
$v\sin i$ (km~s$^{-1}$)                           &28.6 & 8       & 25.3& 8     & ...       &... & 19.2  & 8        & 27.4     & 8  \\
\noalign{\smallskip}
\hline
\end{tabular}
\begin{list}{}{}
\item[$^{\mathrm{a}}$] References: 1 \citet{cohen79}; 
                                   2 \citet{hirth97}, radial velocities (along line of sight) for the 
                                     high-velocity outflow component;
                                   3 \citet{dougados00}, proper motion for DG Tau A;
                                   4 \citet{kenyon95};
                                   5 \citet{white01};
                                   6 \citet{andrews05};
                                   7 \citet{briceno02};
                                   8 \citet{white04};
                                   9 2MASS, \citet{cutri03};
                                   10 \citet{muzerolle98}; 
                                   11 \citet{leinert01};
                                   12 \citet{koresko97};
                                   13 \citet{menard93};
                                   14 \citet{luhman04};
                                   15 \citet{hartigan95};
                                   16 \citet{bacciotti02};
                                   17 \citet{kenyon98};
				   18 \citet{hartmann05};
				   19 \citet{bouvier93}
\item[$^{\mathrm{b}}$] FS = flat-spectrum source (also I/II)  
\item[$^{\mathrm{c}}$] Ages and masses calculated based on parameters given in the cited references, using \citet{siess00} evolutionary tracks 
\item[$^{\mathrm{d}}$] Radius calculated from $T_{\rm eff}$ and $L_{\rm bol}$ 
\item[$^{\mathrm{e}}$] Upper limit to rotation period calculated from radius and $v\sin i$ 
\end{list}
\label{params}
\end{table*}

Basic properties of the four peculiar X-ray sources are collected in Table~\ref{params}. 
The table lists spectral type, infrared 
classification (``YSO class''), T Tauri type according to the H$\alpha$ equivalent width,
the stellar effective temperature $T_{\rm eff}$, the stellar mass based on \citet{siess00} 
evolutionary tracks, the stellar radius computed from the stellar luminosity and $T_{\rm eff}$,
the equivalent width of [OI]~$\lambda 6300$ lines (that can be formed in jets, \citealt{hirth97}), the stellar 
photospheric luminosity $L_*$, the bolometric luminosity $L_{\rm bol}$ (integrated  from 
the infrared+optical spectral energy distribution, including contributions from disks and
envelopes), the extinctions $A_{\rm V}$ and $A_{\rm J}$, the mass accretion rate $\dot{M}_{\rm acc}$,
the mass outflow rate $\dot{M}_{\rm out}$, estimates of the total disk mass $M_{\rm disk}$,
the estimated age as derived from \citet{siess00} evolutionary tracks, the jet radial velocity $v_{\rm rad}$ and
its dispersion $\Delta v_{\rm rad}$, the rotation period (derived from $v\sin i$), and the
projected equatorial velocity $v\sin i$.  For DG Tau, we also list a value for the jet proper motion
$v_{\rm pm}$. Multiple values are given if they are significantly different, as reported in the literature.
A few notes on the individual stars follow.  

{\bf DG Tau A} is a  classical T Tau star whose infrared spectrum, however,
shows a rare, flat infrared spectral energy distribution \citep{andrews05}.
It drives a very energetic jet \citep{mundt83} similar to Class I protostars, including 
a counter-jet (\citealt{lavalley00}). The jets can be  traced out to a distance of 
14.4\arcmin\ \citep{mcgroarty04}. DG Tau is one of the most active CTTS known, 
and ranks among the CTTS with the  highest mass accretion and mass outflow rates, competing with 
well-studied protostellar jet sources \citep{hartigan95, bacciotti02}. The star has indeed been 
suggested to be a transition object between protostars and CTTS \citep{pyo03}. On the other hand, 
the stellar visual extinction is rather modest, with reported $A_{\rm V}$ in the range of 
1.4--3.3~mag \citep{hartigan95, muzerolle98, white01, white04}. A much  lower visual extinction
has been determined for the jet's Herbig-Haro objects, namely $A_{\rm V} = 0.39$~mag \citep{cohen85}.

The jet is also known from radio
observations \citep{cohen86}. High-resolution studies have shown that a  narrow jet ploughs 
through progressively slower and wider outflow structures.
A few arcseconds from the star, the proper motion of the jet indicates velocities in the plane of the sky
of $194\pm 20$~km~s$^{-1}$, on average \citep{dougados00}, with a maximum of 
$> 360$~km~s$^{-1}$ \citep{eisloeffel98}. The jet radial velocity is 210--250~km~s$^{-1}$, compatible
with the reported inclination angle of  38~deg \citep{eisloeffel98}. Bulk gas speeds reach  500~km~s$^{-1}$ 
\citep{bacciotti00, beristain01}, with FWHM line widths of 100--200~km~s$^{-1}$ \citep{pyo03}.

The high-velocity jet shows bow-shock like structures out to distances of several arcsecs 
(also known as HH~158, \citealt{mundt83}) but can be followed down to 0.1\arcsec\  from the star 
\citep{bacciotti00, pyo03}.  Recently, \citet{bacciotti02}
measured rotation of the jet around its flow axis, which has subsequently been 
used to infer the origin of the jet in the inner disk (0.3$-$4~AU for the lower-velocity 
component, and possibly in the X-wind region at the inner-disk edge at $~0.1$~AU for the 
high-velocity jet;  \citealt{anderson03, pyo03}). The jet mass-loss rate has been estimated
by various authors in the range of   $(2.4-7.9)\times 10^{-7}M_{\odot}$yr$^{-1}$ \citep{hartigan95,
muzerolle98, bacciotti02, white04}. This rate amounts to about 30\% of the accretion rate, 
$\dot{M}_{\rm acc} = (0.8-2)\times 10^{-6}M_{\odot}$yr$^{-1}$ (\citealt{hartigan95, white04}; \citealt{white01}
give a somewhat lower accretion rate).
Shocks  seem to be the principal source of  heating and excitation \citep{lavalley00}. We note that DG Tau B,
a jet-driving protostar, is located at a considerable distance  from  DG Tau A (separation 
$\approx 50\arcsec$), much larger than the PSF of the EPIC cameras. We will discuss this second
source further below.

{\bf GV Tau = Haro 6-10 AB} is a close, embedded binary system with a separation of 1.3\arcsec, consisting of an
optically revealed T Tau-like star and a deeply embedded ``infrared companion''   considered to be
as a protostar. The infrared spectrum of the system is that of a protostar (Class I, Table~\ref{params}).
This binary has  received  particularly detailed  study in the radio range
\citep{reipurth04}. The jet shows radial velocities in the range 80--200~km~s$^{-1}$ \citep{hirth97}.

{\bf DP Tau} drives a jet that has been recorded to a distance of at least 27$\arcsec$, and possibly
to 111$\arcsec$. The jet is oriented  relatively close to the plane of the sky \citep{mundt98} 
although radial velocities of the jet still reach about 110~km~s$^{-1}$ \citep{hirth97}. The star
shows strong [OI] emission from the jets \Citep{hirth97}. 

{\bf CW Tau} drives a bipolar outflow that can be traced out to at least 4--6$\arcsec$ from the star \citep{hirth94},
with Herbig-Haro objects found out to  16.8\arcmin\ \citep{mcgroarty04}. 
A dedicated study of this jet was presented by the latter authors. \citet{dougados00} reported a highly collimated, 
transversally resolved jet structure.

\subsection{Jet-driving protostars}

We took advantage of the high spatial resolution of {\it Chandra}  to obtain images of four 
protostellar sources that drive strong, collimated jets and outflows. Their circumstellar
disks are seen nearly edge-on so that any soft emission will be strongly absorbed {\it unless}
it is originating sufficiently far away from the star. This allows us to distinguish between  
X-ray sources coincident with the protostars and softer X-rays from internal shocks in the  
jets \citep{bally03} and will thus assist in our interpretation of the soft components seen
in the T Tau sources. This geometry also permits 
us to probe the disks and envelopes themselves, by studying photoelectric absorption.

Two objects, \object{DG Tau B} and \object{FS Tau B} = \object{Haro 6-5B}, are part
of the XEST survey, but we also discuss here two further stars that are located in the Orion region
at larger distances, namely the protostellar sources of the \object{HH~34} and the \object{HH~111} jets. 

We used the {\sl wavdetect} algorithm included in the {\it Chandra} CIAO  software
to identify relevant X-ray sources and to determine their positions. The algorithm was
applied to images containing photons in the 0.3--10~keV range, 
the inspected field containing approximately 
$10^6$ standard ACIS pixels of size $0.5\arcsec\times 0.5\arcsec$.
We set the  {\sl sigthresh} parameter at a value of $10^{-6}$, which means that
only one X-ray detection in the field is likely to be spurious.

To optimize the positional information for the X-ray sources, we  registered, as far as possible, 
field sources against 2MASS catalog entries \citep{cutri03} but found that the positional agreement  
is extremely good, with typical offsets\footnote{The 90\% source location error circle
in {\it Chandra} has a radius of about 0.5\arcsec, see {\it Chandra} Proposes' Observatory Guide v.8.} 
of $<0.4^{\prime\prime}$, which is  not significant.  We provide some specific information on the four sources below.

{\bf DG Tau B}  is  an embedded, envelope-dominated, jet-driving protostar  described in
detail by, among others, \citet{mundt83}, \citet{mundt87}, and  \citet{eisloeffel98}.
Near-infrared imaging shows two narrow cavities and a thick dust-lane seen nearly edge-on, but probably not the
protostar (\citealt{padgett99} = P99). 
DG Tau B is considered to be in a Class I 
stage \citep{watson04}. It was detected as a radio source by \citet{rodriguez95b}.

{\bf FS Tau B = Haro 6-5B} is a jet-driving embedded source about 20\arcsec\ west of the classical
T Tau binary system FS Tau A.  It was discovered as a jet source by \citet{mundt84} and described
in detail by \citet{mundt87}, \citet{eisloeffel98}, and \citet{krist98}.
Near-infrared imaging shows it as a disk-dominated protostar in which a thick dust lane, seen nearly edge-on, 
separates two nebulae considered to be the illuminated surfaces  of  the flared disk, with a widely 
open cavity (P99). FS Tau B is probably in a rather evolved stage of Class I evolution,  in
transition to becoming a TTS. In the near-infrared, the star itself seems to be detected (P99). 
FS Tau B was found as a radio source by \citet{brown85}.

\begin{figure*}[t!]
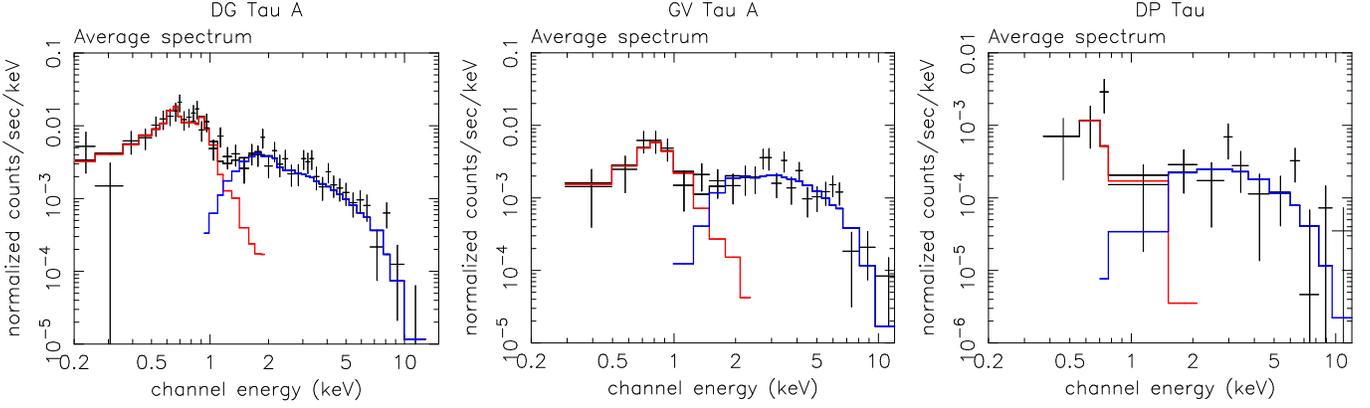

\hbox{
\includegraphics[angle=-90,width=5.8cm]{guedel_f1a.ps}
\hskip 0.2truecm\includegraphics[angle=-90,width=5.8cm]{guedel_f1b.ps}
\hskip 0.2truecm\includegraphics[angle=-90,width=5.8cm]{guedel_f1c.ps}
}
\caption{Average spectra of DG Tau A (left), GV Tau (middle), and DP Tau (right). Also shown are 
the fits to the spectra (black histograms) and separately the soft 
(red) and the hard (blue) spectral components. }
\label{specave} 
\end{figure*}

The {\bf HH~34} source drives one of the finest, highly collimated jets \citep{buehrke88}. The optical jet can 
be followed to within $1^{\prime\prime}$ of the Class I protostar. The star itself  has been detected
in the optical \citep{buehrke88}, in the infrared (IR) \citep{reipurth02}, and in the  radio \citep{rodriguez96}, 
with a  molecular cloud surrounding it.

The {\bf HH~111} infrared source drives another highly collimated, long jet but the source
and the innermost 22\arcsec\ of the  jet are very highly  obscured \citep{reipurth97}. 
The absorbed portions of the jet have  been revealed in the infrared by \citet{reipurth99} and 
\citet{reipurth00}. Radio detections have also been reported by 
\citet{rodriguez94} and \citet{reipurth99}. They identify a quadrupolar outflow, suggesting 
that the driving source is a close binary.

\section{Results}\label{sect:results}

\subsection{Optically revealed, jet-driving T Tau stars}

\subsubsection{Spectral interpretation}

Fig.~\ref{specave} presents EPIC PN spectra from DG Tau A, GV Tau, and DP Tau, respectively,
together with spectral fits (see below). We have used the full exposure times except for a few intervals 
with high background radiation. The MOS spectra are similar but considerably fainter. They were used
for spectral fits together with the PN spectra in the case of DG Tau A, while they were
too faint to add useful information in the other cases.

All three spectra are rather flat and show two shallow maxima, one around  0.65--0.8~keV and the
other at 1.5--3~keV, with an intervening trough at 1--1.5~keV. These spectra cannot be  acceptably
fitted with a combination of thermal coronal models subject to a {\it single} photoelectric
absorption component, as was already pointed out by \citet{guedel05} for the {\it Chandra} ACIS
spectrum of DG Tau A. The softer maximum is formed by emission lines of O\,{\sc viii} Ly$\alpha$ at 
18.97~\AA\ (0.65~keV), Fe\,{\sc xvii} at 17~\AA\ (0.73~keV)
and at 15~\AA\ (0.83~keV), as well as Fe\,{\sc xviii} at 16~\AA\ (0.78~keV, blended with O\,{\sc viii} Ly$\beta$) and
at 14.2~\AA\ (0.87~keV). The Ne\,{\sc ix}  He-like triplet contributes at $\approx 13.5$~\AA\ (0.92~keV, 
blended with F\,{\sc xvii} at 13.8~\AA\ or 0.9~keV). All these lines have maximum formation temperatures 
$T_{\rm max}$ between 3.2~MK and 6.9~MK (O\,{\sc viii}: 3.2 MK; Ne\,{\sc ix}: 3.9~MK; Fe\,{\sc xvii}: 5.2~MK; Fe\,{\sc xviii}: 
6.9~MK). The best matches of the subpeaks at 0.7~keV and at 0.80--0.85~keV in the
DG Tau A spectrum are clearly the O\,{\sc viii} Ly$\alpha$ and the two Fe\,{\sc xvii} lines. On the other hand, 
we note the absence of strong emission around 1~keV usually present in very active stars 
with hot coronae \citep{telleschi05}. 
This latter emission would be formed by Ne\,{\sc x} at 12.1~\AA\  (1.0~keV, $T_{\rm max} = 5.9$~MK), Fe\,{\sc xx} 
at 12.8~\AA\ (0.97~keV), and by various further lines of Fe\,{\sc xx-xxiv} ($T_{\rm max} \approx 12-20$~MK). For typical
elemental compositions, these spectral properties clearly suggest emission by a rather cool plasma 
{\it in the absence of prominent contributions from hot components}. The rather high emission level down
to 0.2~keV in DG Tau A  and to $\approx 0.3-0.4$~keV in DP Tau further suggests a rather low absorbing column density.

On the other hand, the spectra above $\approx 1.5$~keV are  shallow and can, despite the low
count rates, be followed out to 10~keV for all three stars. This unambiguously indicates emission
from very hot ($> 10$~MK) plasma which would normally form strong emission lines around 1~keV
as well. The hard emission is dominated by
bremsstrahlung which, in the absence of absorption,  rises toward lower energies. But these low-energy
features of hot plasma are suppressed by strong photoelectric absorption: All three hard spectral
components reveal a turnover around 2~keV.

As suggested by \citet{guedel05}, these spectra can be  successfully fitted with a model consisting
of two thermal components, each subject to a {\it separate} photoelectric absorption component.  
We fitted the spectra as follows. For DG Tau A and GV Tau, we first considered counts within the 
soft maximum only (0.25--1.0~keV). The abundances adopted (and held fixed)
reflect an ``inverse First Ionization Potential'' distribution often found in magnetically
active and pre-main sequence stars (see \citealt{guedel06} for a discussion). Fit parameters were
the absorbing hydrogen column density $N_{\rm H,s}$, a single temperature $T_{\rm s}$, and the emission measure EM$_{\rm s}$.
The fits were perfect, with reduced $\chi^2\la 1$. Next, we added the hard portion of the observed spectrum
up to 10~keV but held the fit obtained above fixed. We added a second spectral component with
its own absorption component and fitted the corresponding variables $N_{\rm H,h}$, $T_{\rm h}$, and EM$_{\rm h}$. To optimize in 
particular the overlapping spectral region (1--1.5~keV), we continued fitting the $N_{\rm H}$, $T$, and EM 
parameters of both components simultaneously, but this did not significantly change the parameters. 
The DP Tau spectrum is too poor to allow for step-wise fitting. We therefore fitted the two components
simultaneously from the outset.

\begin{table*}
      \caption[]{X-ray properties of the TAX sources}
         \label{Xparams}
         \begin{tabular}{llllllll}
            \hline
            \noalign{\smallskip}
                      &  \multicolumn{3}{c}{\rm Soft\ component}        &      \multicolumn{3}{c}{\rm Average\ hard\ component}        &\\
                      &  \multicolumn{3}{c}{\hrulefill}                 &      \multicolumn{3}{c}{\hrulefill}                                            &      \\
             \noalign{\smallskip}
              Star         &  $N_{\rm H,s}^{a}$                 &  $T_{\rm s}^{a}$   &  $L_{\rm X,s}^{b}$         &   $N_{\rm H,h}^{a}$             &   $T_{\rm h}^{a}$     &   $L_{\rm X,h}^{b}$                      & $\chi^2/{\rm dof}$   \\
                           &   $(10^{22}~{\rm cm}^{-2})$    &  $(10^{6}~{\rm K})$ &   $(10^{29}{\rm ~erg~s}^{-1})$ &   $(10^{22}~{\rm cm}^{-2})$ & $(10^{6}~{\rm K})$    &   $(10^{29}{\rm ~erg~s}^{-1})$       \\
             \noalign{\smallskip}
            \hline
             \noalign{\smallskip}
              DG Tau A$^c$  &  0.11\ (0.08-0.14)           &  3.7\ (3.2-4.6)   &  0.96                      &  1.80\ (1.21-2.40)         &  69\ (44-188)        &  5.1      & 89.9/93             \\
                            &  = 0.45                      &  2.2\ (2.05-2.35) &  9.6                       &  1.71\ (1.14-2.30)         &  73\ (45-220)        &  5.0      & 100.2/94             \\
             \noalign{\smallskip}
            \hline
             \noalign{\smallskip}
             GV Tau A       &  0.12\ (0.01-0.79)           &  5.8\ (2.1-9.1)   &  0.54                      &  4.12\ (2.07-6.47)         &  80\ (34-...)        &  10.2     & 20.3/19           \\
                            &  = 1.12                      &  0.94\ (0.94-1.31)&  4170                      &  3.48\ (1.69-5.48)         &  95\ (42-...)        &  9.6      & 24.6/20           \\
             \noalign{\smallskip}
            \hline
             \noalign{\smallskip}
              DP Tau        &  0.00\ (0.00-1.05)           &  3.2\ (0.0-7.7)   &  0.04                      &  3.78\ (1.38-14.0)         &  61\ (10-...)        &  1.1      & 8.9/7            \\
                            &  = 0.29                      &  2.3\ (0.94-3.8)  &  1.1                       &  3.13\ (0.78-13.8)         &  104\ (10-...)       &  0.92     & 9.1/8              \\
            \noalign{\smallskip}
            \hline
         \end{tabular}
  \begin{list}{}{}
  \item[$^{\mathrm{a}}$] 90\% confidence intervals in parentheses (ellipses indicating unconstrained parameter)
  \item[$^{\mathrm{b}}$] Modeled for the 0.1--10~keV energy interval
  \item[$^{\mathrm{c}}$] Fits applied to combined PN, MOS1, and MOS2 data
  \end{list}
\end{table*}

As far as abundances are concerned, the spectrum of DG Tau
marginally suggested a somewhat higher Fe abundance (0.4 instead of the adopted 0.2 times the
solar photospheric value given by \citealt{anders89}), but this value was poorly constrained,
and the other parameters of interest here, $kT$ and $N_{\rm H}$, did not change in any significant way.
We further experimented with variable abundances as follows. We kept the adopted abundance
ratios fixed but varied the absolute abundance level. For DG Tau A, the best fit resulted in unacceptably
low abundances of Fe ($< 1$\% of the solar value), while the temperatures and absorption column
densities again remained similar.  

The final results are reported in Table~\ref{Xparams}. 
The X-ray luminosities $L_{\rm X,s}$ and $L_{\rm X,h}$  refer to the 0.1--10~keV range based on 
integration of the best-fit model, and assume a distance of 140~pc of the Taurus association
(e.g., \citealt{loinard05, kenyon94}).

Noting that the $N_{\rm H}$ value for the soft component of DG Tau A are lower than those anticipated 
from optical extinction, and $N_{\rm H}$ is only marginally in agreement with the
lowest $A_{\rm V}$ reported for GV Tau A, we next adopted a fixed $N_{\rm H}$ value for the soft component (marked by ``='' in Table~\ref{Xparams}),
using the standard conversion formula, $N_{\rm H} \approx 2\times 10^{21}A_{\rm V}$~cm$^{-2}$, 
valid for standard gas-to-dust ratios (see \citealt{vuong03} and references therein). These fit results are also reported in 
Table~\ref{Xparams}.
Fixing $N_{\rm H}$ at the expected values provides significantly worse  
fits and lowers the temperature of the cooler component, at the same time rising its
$L_{\rm X}$.  The value of the latter is unacceptable for GV Tau because it approaches  $L_*$,
while $L_{\rm X}$ is usually bound by $L_{\rm X}/L_* \la 10^{-3}$ (\citealt{guedel04} and references therein). 
We believe that the origin of this excess is 
purely numerical: increased $N_{\rm H}$ allows more and cooler plasma to be present without 
significant spectral contribution, because X-rays from cooler plasma are softer and therefore
more strongly absorbed. The effect continues as one increases $N_{\rm H}$. This lends
most credibility to the best fit without enforced $N_{\rm H}$, and this best fit indeed results in only modest
suppression even at low energies. 

In summary, the spectra of DG Tau A, GV Tau, and DP Tau are thus composed of two components, a cool component
subject to very low absorption and a hot component subject to photoelectric absorption
about one order of magnitude higher. The cool component shows temperatures atypical for T Tau stars \citep{skinner03, telleschi06},
ranging from 3--6~MK (best fit), while the hot component reveals extremely high temperatures
(60--100~MK). If $N_{\rm H}$ is fitted to the spectrum, it tends to be lower than expected
from the optical extinction.

The spectrum of CW Tau resembles the spectrum of DP Tau but is too weak to derive definitive results. This source is also
affected by the wings of the nearby and very bright X-ray source of V773~Tau (at a distance of 1.5\arcmin). 
A single-$T$ fit to the harder photons (above 1.2~keV)
requires an excessively absorbed ($N_{\rm H} \approx 10^{23}$~cm$^{-2}$), very strong cool component  with $T \approx 4$~MK,  
but there is a marginal excess of counts below 1~MK that requires a further, low-absorption component. Better-quality
data are needed. We  note that the star's $A_{\rm V} = 2.29$~mag \citep{kenyon95}  suggests  a much smaller
$N_{\rm H} \approx 5\times 10^{21}$~cm$^{-2}$.

Results for the three further jet/outflow driving sources mentioned above, UZ Tau, DD Tau, and XZ Tau, are
summarized in \citet{guedel06}.  UZ Tau (a triple system) reveals a relatively soft spectrum (emission-measure
weighted average temperature of 5--7~MK). On the  contrary, the very faint spectrum of DD Tau is extremely hard, reminiscent
of the harder component  of the sources discussed above, with $T_{\rm av} = 38-42$~MK and excess absorption 
($N_{\rm H} = [2.3-2.9]\times 10^{21}$~cm$^{-2}$, whereas $A_{\rm V} = 0.39$~mag after \citealt{white01}, or $A_{\rm J} = 0.05$~mag after \citealt{briceno02}). 
A soft excess is not found for this source.  Adopting a jet bulk velocity of 300~km~s$^{-1}$ and a radial velocity of
$v_{\rm rad} \approx 75$~km~s$^{-1}$ \citep{hirth97}, we infer an inclination angle of $\approx$75~deg. 
The strong photoelectric absorption may be due to disk material, while the low values of $A_{\rm V}$ 
and $A_{\rm J}$ may be affected by scattered light. No anomaly was found for XZ Tau. 
 
\subsubsection{Light curves and time-resolved spectroscopy}

\begin{figure}[t!]
\includegraphics[angle=-0,width=7.2cm]{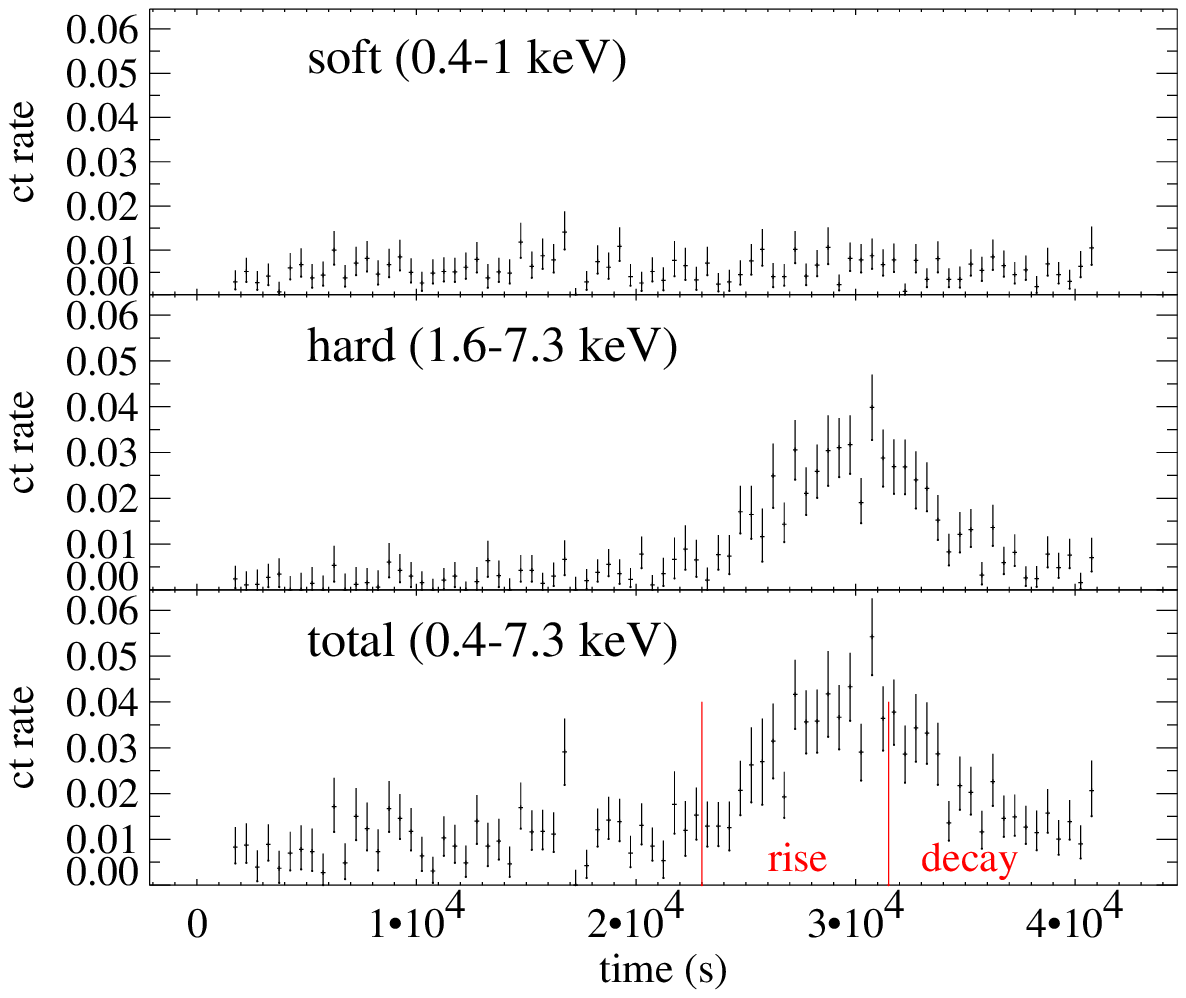}
\vskip 0.2truecm
\includegraphics[angle=-0,width=7.2cm]{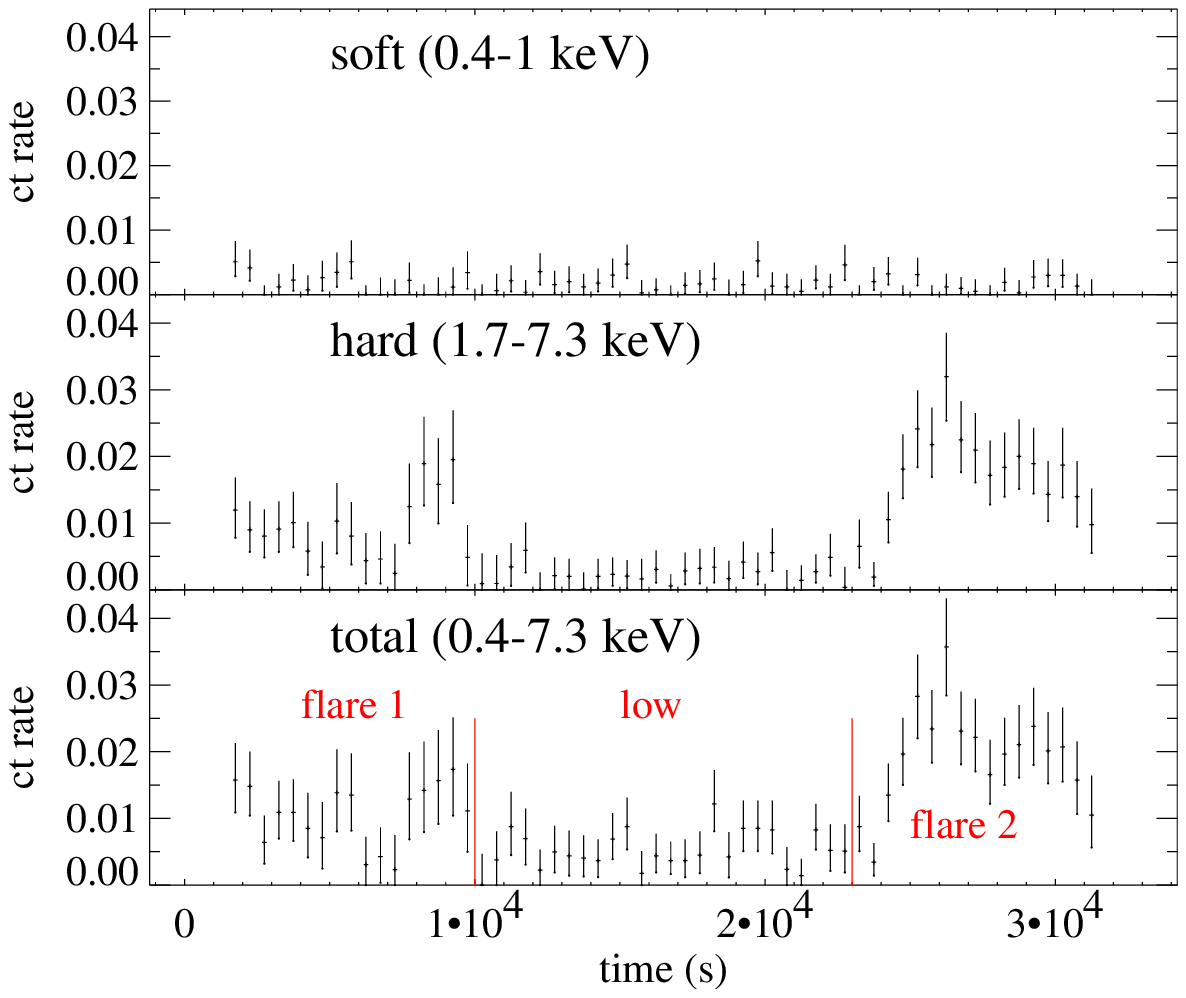}
\vskip 0.2truecm
\includegraphics[angle=-0,width=7.2cm]{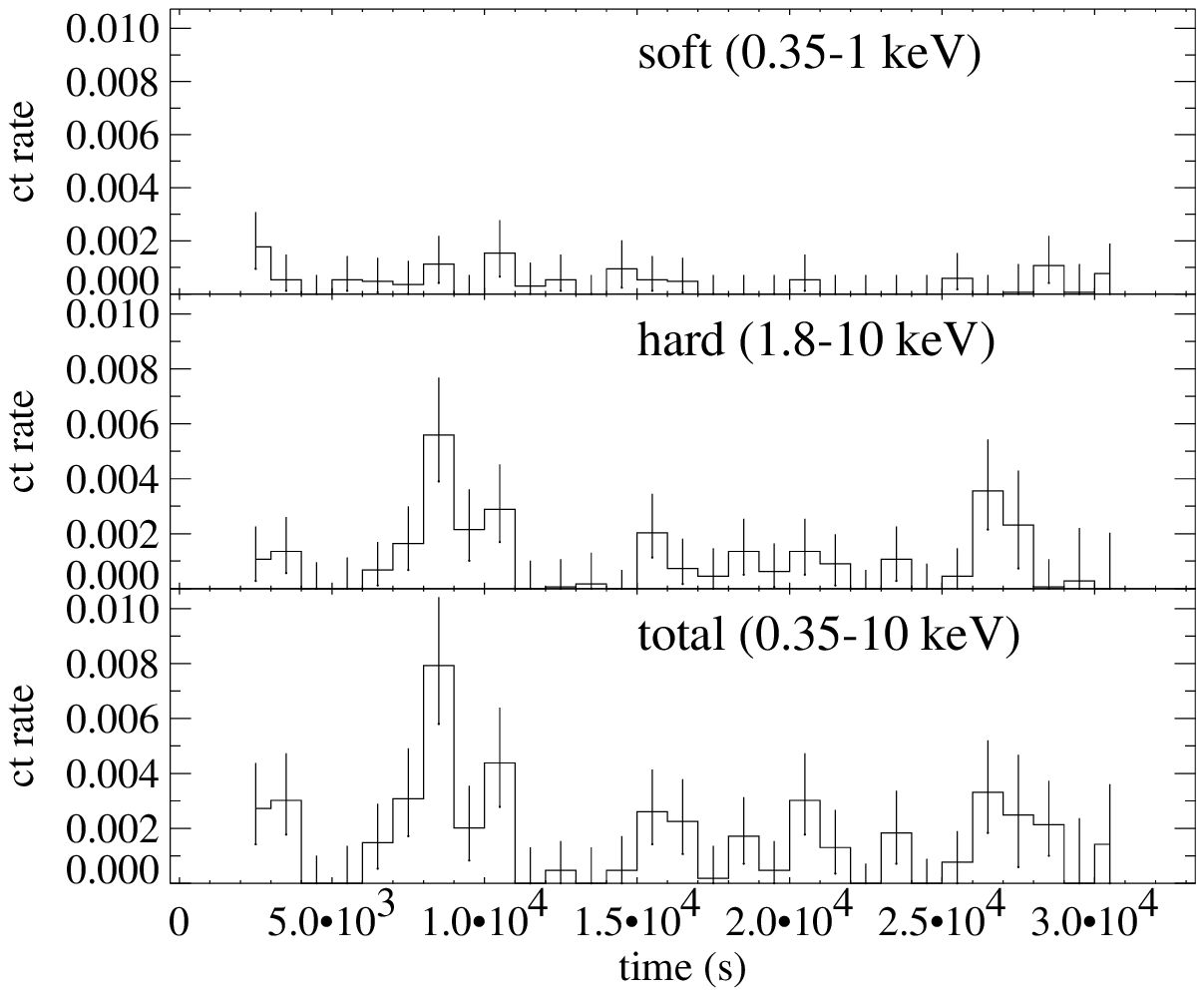}
\caption{Background-subtracted light curves of DG Tau A (top), GV Tau (middle), and DP Tau (bottom),
 binned to 500~s, 500~s, and 1000~s, respectively. Each figure
 contains three panels, showing, from top to bottom:   
   Soft photons (0.4--1.0~keV for DG Tau A and GV Tau, and 0.35--1~keV for DP Tau); 
   hard photons (1.6--7.3~keV for DG Tau A, 1.7--7.3~keV for GV Tau, 1.8--10~keV for DP Tau); 
   full band	(0.4--7.3~keV for DG Tau A and GV Tau, and 0.35--10~keV for DP Tau).  For 
   DG Tau A and GV Tau, intervals separately studied are marked by red vertical lines. Only the time intervals
   observed jointly by all three EPIC cameras are shown.}
\label{light}%
\end{figure}

We present the light curves in Fig.~\ref{light}, where we used counts below $\approx 1$~keV and above $\approx 1.6-1.8$~keV
(depending on the spectral shape)
for the soft and the hard component, respectively. All light curves show clear signatures of 
flaring. Surprisingly, however, flaring is revealed only in the hard component while the soft component remains 
at a low level. We tested the background-subtracted light curves for constancy, applying a $\chi^2$ test. To do  so, we 
rebinned the light curve of DG Tau A such that the bins of the soft curve contained, on average, 8  or 11 counts.
This required bins of 1500-2000 seconds, which is sufficient to still recognize typical flares.
With the same binning, the hard curve contained, on average, 6 or 8 cts but showed the strong flare well resolved.
For GV Tau, we binned to 2000 and 3000~s, which on average resulted in only 2.3 and 5~cts per bin, respectively.
The hard curve shows a 2.7 times higher count rate. 
A test for constancy resulted in a  reduced $\chi^2$ that was smaller than unity  
for the soft light curves of DG Tau A and GV Tau, whereas the  variability in the hard
curve is highly significant with $\chi^2_{\rm red} \approx 10-17$.  
 DP Tau is too faint for a meaningful $\chi^2$ test of this kind.
 
\begin{figure*}
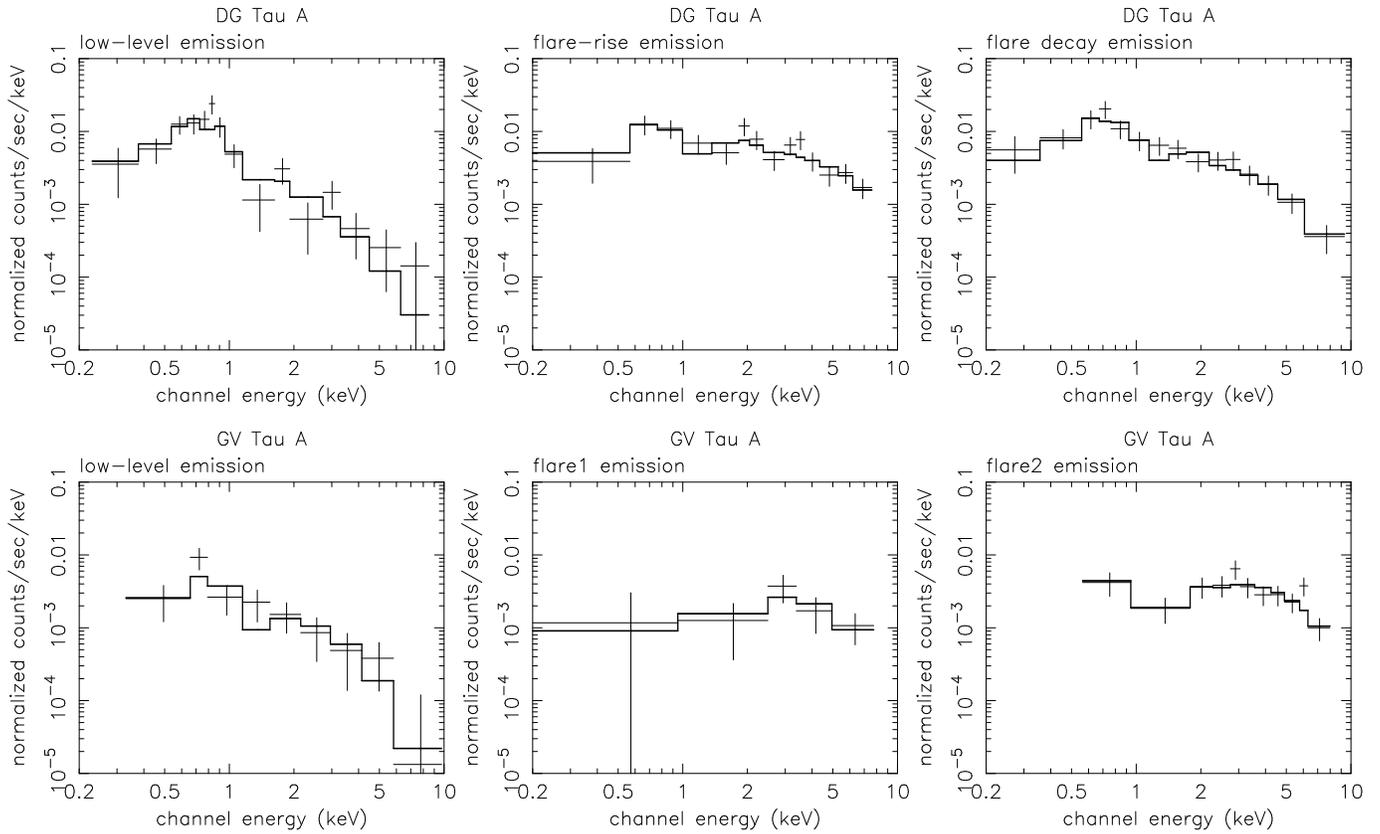

   \hbox{
   \includegraphics[angle=-90,width=5.95cm]{guedel_f3a.ps}
   \includegraphics[angle=-90,width=5.95cm]{guedel_f3b.ps}
   \includegraphics[angle=-90,width=5.95cm]{guedel_f3c.ps}
   }
   \vskip 0.3truecm
   \hbox{
   \includegraphics[angle=-90,width=5.95cm]{guedel_f3d.ps}
   \includegraphics[angle=-90,width=5.95cm]{guedel_f3e.ps}
   \includegraphics[angle=-90,width=5.95cm]{guedel_f3f.ps}
   }
   \caption{{\it Upper panel:} Spectra and fits of DG Tau A during low-level episodes (left),
   during the rise to the large flare (middle) and during the decay phase (right).
   {\it Lower panel:} Spectra and fits of GV Tau during low-level episodes (left),
   during the first flare episode (middle) and during the second flare episode (right).}
    \label{DGTau}%
    \end{figure*}

To  further clarify the spectral behavior in time, we extracted EPIC spectra during various episodes
for DG Tau A and GV Tau (low level, flare rise, and flare decay for DG Tau A; low level,
first flare episode and second flare episode for GV Tau; see Fig.~\ref{light} for a graphical definition of the 
corresponding time intervals). The corresponding plots are shown in 
Fig.~\ref{DGTau}. The hard component varies both in flux
and in spectral shape: the shallower slope during the flares  indicates higher temperatures, while
the soft component remains at the same level. This is borne out by spectral fits performed
for the same intervals. The results are reported in Table~\ref{Xvariab}. Because the spectra are
too poor to fit $N_{\rm H}$ reliably, this parameter was frozen at the value found for  each integrated
spectrum. The temperature of the soft component remains constant within the errors, while the temperature
of the hard component rises during the flare episodes. Note also that the X-ray luminosities of
the soft component are very similar (the spectrum of the first flare episode of GV Tau is too faint
to produce reliable results).

\begin{table*}
      \caption{Time-dependent X-ray properties of the TAX sources}
         \label{Xvariab}
         \begin{tabular}{llllllll}
            \hline
            \noalign{\smallskip}
                      &  \multicolumn{3}{c}{\rm Soft\ component}         &      \multicolumn{3}{c}{\rm Average\ hard\ component}               & \\
                      &  \multicolumn{3}{c}{\hrulefill}            &      \multicolumn{3}{c}{\hrulefill}                                            &      \\
             \noalign{\smallskip}
           {\rm Star}      &  $N_{\rm H,s}^{a}$                    &  $T_{\rm s}^{a}$    & $L_{\rm X,s}^{b}$               &   $N_{\rm H,h}^{a}$             &   $T_{\rm h}^{a}$     &   $L_{\rm X,h}^{b} $                 & $\chi^2/{\rm dof}$   \\
                           &   $(10^{22}~{\rm cm}^{-2})$         &  $(10^{6}~{\rm K})$ &   $(10^{29}{\rm ~erg~s}^{-1})$ &   $(10^{22}~{\rm cm}^{-2})$ & $(10^{6}~{\rm K})$    &   $(10^{29}{\rm ~erg~s}^{-1})$       \\
             \noalign{\smallskip}
            \hline
             \noalign{\smallskip}
            DG Tau A low$^c$	& =:0.11  		  &  3.7\ (3.3-4.6)   &  0.91			   &  =:1.80		       &  23\  (14-42)        &  2.1			      & 24.5/33 	    \\
            DG Tau A rise$^c$	& =:0.11  		  &  3.9\ (3.1-5.6)   &  0.89			   &  =:1.80		       &  396\ (144-...)      &  12.4			      & 27.9/31 	    \\
            DG Tau A decay$^c$  & =:0.11  		  &  3.9\ (3.3-4.9)   &  1.02			   &  =:1.80		       &  74\  (44-144)       &  6.7			      & 15.7/24 	    \\
             \noalign{\smallskip}
            \hline
             \noalign{\smallskip}
            GV Tau  low       &  =:0.12 		  &  4.7\ (3.0-9.7)   &  0.54			   &  :=4.12		       &  17.5\ (4.3-63)       &  7.3			      & 5.4/5		  \\
            GV Tau  flare 1   &  =:0.12 		  &  2.1\ (...)       &  0.16			   &  =:4.12		       &   77\ (24-...)        &  16.3  		      & 3.7/2	       \\
            GV Tau  flare 2   &  =:0.12 		  &  4.5\ (...-12.6)  &  0.45			   &  =:4.12		       &  119\ (52-...)        &  19.9  		      & 6.7/7	       \\
            \noalign{\smallskip}
            \hline
         \end{tabular}
  \begin{list}{}{}
  \item[$^{\mathrm{a}}$] 90\% confidence intervals in parentheses (ellipses indicating unconstrained parameter)
  \item[$^{\mathrm{b}}$] Modeled for the 0.1--10~keV energy interval
  \item[$^{\mathrm{c}}$] Fits applied to combined PN, MOS1, and MOS2 data
  \end{list}
\end{table*}

\subsubsection{Comparison with Chandra spectra}\label{longterm}

\begin{figure}[t!]
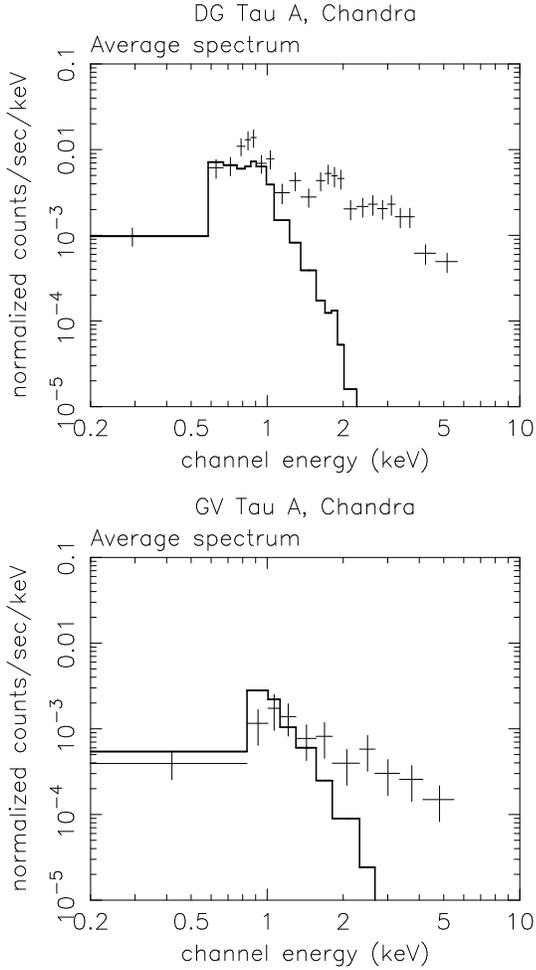

   \centering
   \includegraphics[angle=-90,width=7cm]{guedel_f4a.ps}
   \vskip 0.3truecm
   \includegraphics[angle=-90,width=7.cm]{guedel_f4b.ps}
   \caption{{\it Chandra}  spectrum of DG Tau A (from ACIS-S) and GV Tau (from ACIS-I). 
            The histogram  corresponds to the soft component modeled from {\it XMM-Newton} 
	    data but folded through the appropriate ACIS responses.}
              \label{GVTauChandraSpec}%
    \end{figure}

Both DG Tau A and GV Tau have been observed by {\it Chandra} ACIS as well (see \citealt{guedel06} for 
an observation log). The sources are rather faint, and no flaring was detected during the
observations. We therefore present only the integrated spectra and compare them with the integrated
{\it XMM-Newton} spectra. Figure~\ref{GVTauChandraSpec} shows the observed {\it Chandra} spectra together
with the spectral model of the {\it soft} component of the {\it XMM-Newton} observations, folded with
the appropriate ACIS response matrices. We note that the {\it Chandra} and {\it XMM-Newton} observations
were separated by 7 months for DG Tau A and by 8 months for GV Tau.

Whereas the  flux of the hard components  varied greatly during the 30--40~ks {\it XMM-Newton} 
observations, the flux of the soft components changed little in the 7--8 months between the 
{\it Chandra} and {\it XMM-Newton} observations, although some details in the 
spectrum do differ. The {\it Chandra} spectrum of DG Tau A shows excess flux at 0.8--0.9~keV,
a discrepancy that did not disappear if the the soft portions ($<$1~keV) of both spectra were fitted 
simultaneously, even if we changed the abundances.  Fitting the {\it Chandra} spectrum
finds a best-fit Fe abundance of $\approx 0.2$ \citep{anders89}, the same that we adopted for the {\it XMM-Newton}
model. The soft-component $N_{\rm H}$ is again low ($\approx 0.6\times 10^{21}$~cm$^{-2}$) and marginally compatible 
with {\it XMM-Newton},
although the best-fit temperature is somewhat higher (4.4~MK).  No significant deviation between the
soft {\it Chandra} and {\it XMM-Newton} spectra is found for GV Tau.

\subsection{Embedded, jet-driving protostars}

The embedded, jet-driving protostars seen at high inclination expectedly reveal rather
faint X-ray fluxes owing to strong photoelectric absorption. Table~\ref{proto} summarizes X-ray detection
properties and results from spectral fits (see below).

\begin{figure}
\resizebox{0.8\hsize}{!}{\includegraphics{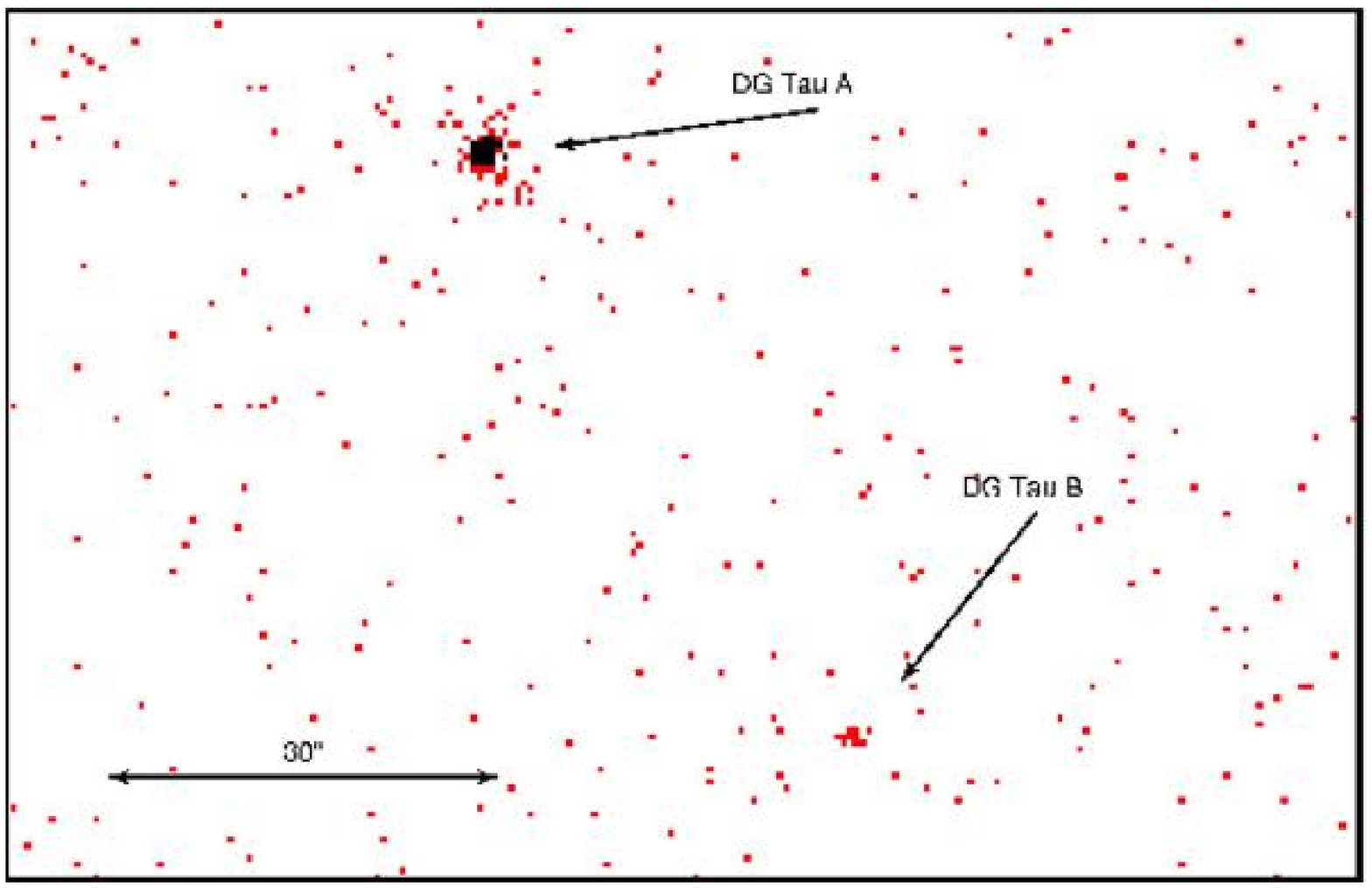}}
\resizebox{0.8\hsize}{!}{\includegraphics{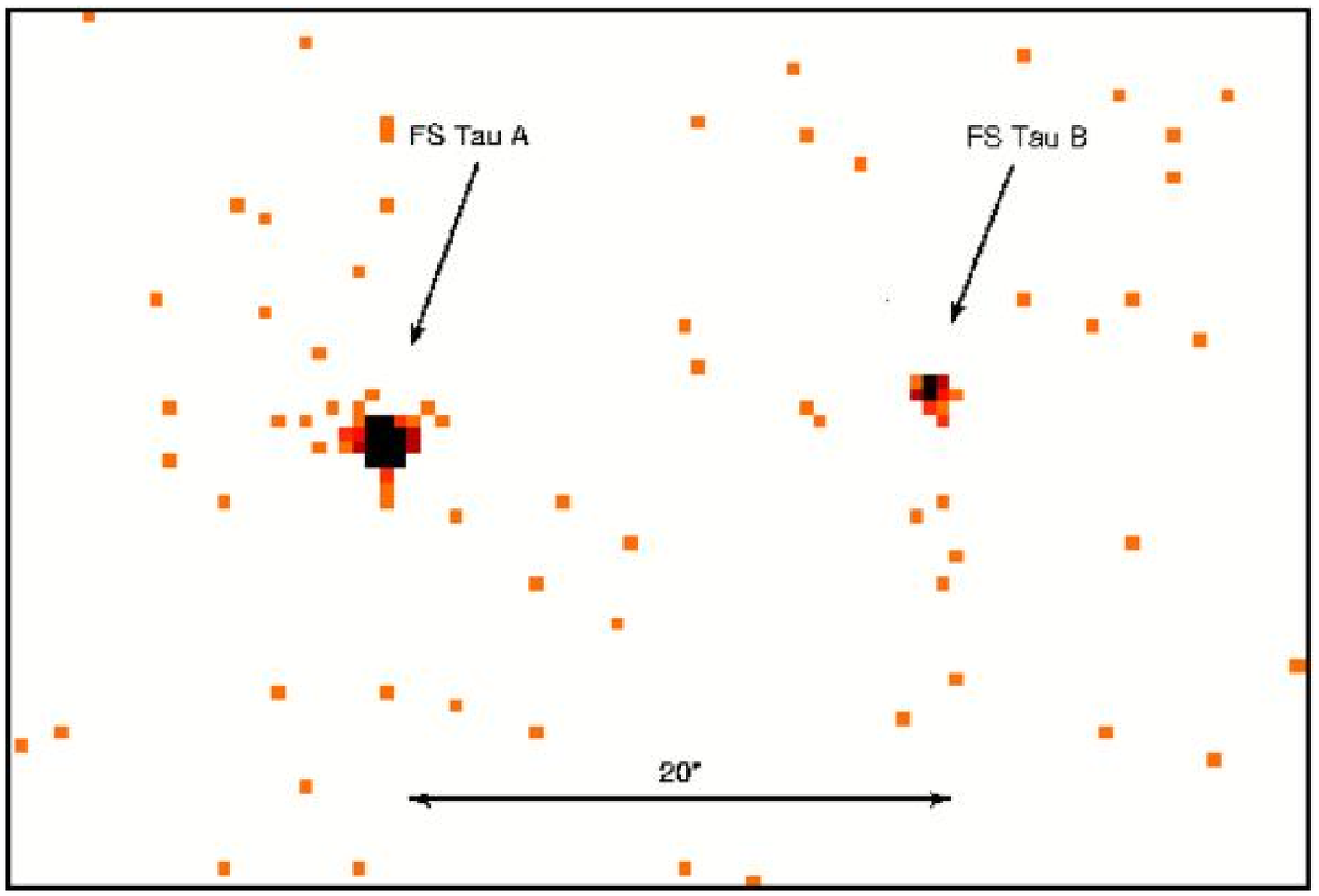}}
\resizebox{0.8\hsize}{!}{\includegraphics{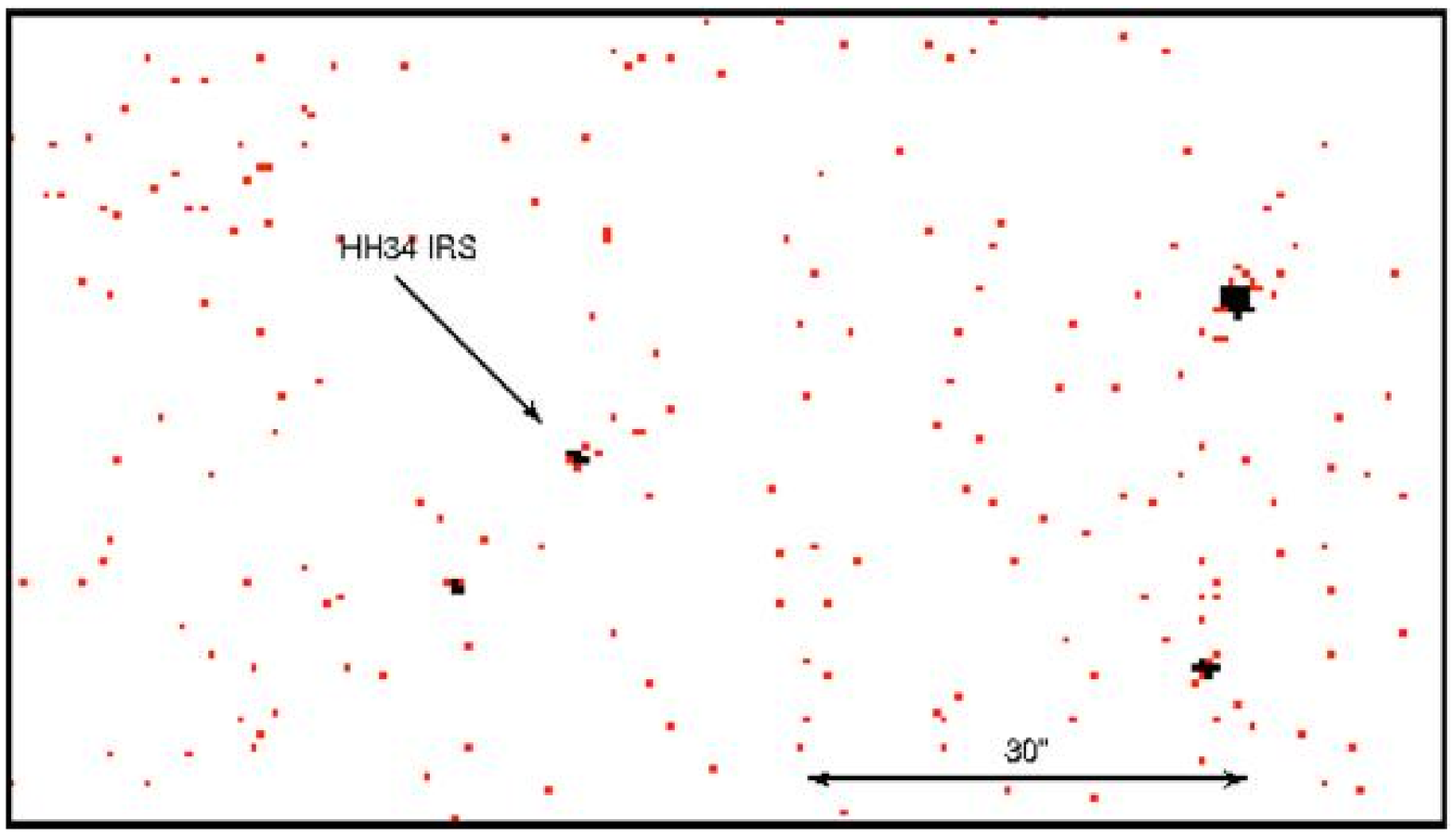}}
\resizebox{0.8\hsize}{!}{\includegraphics{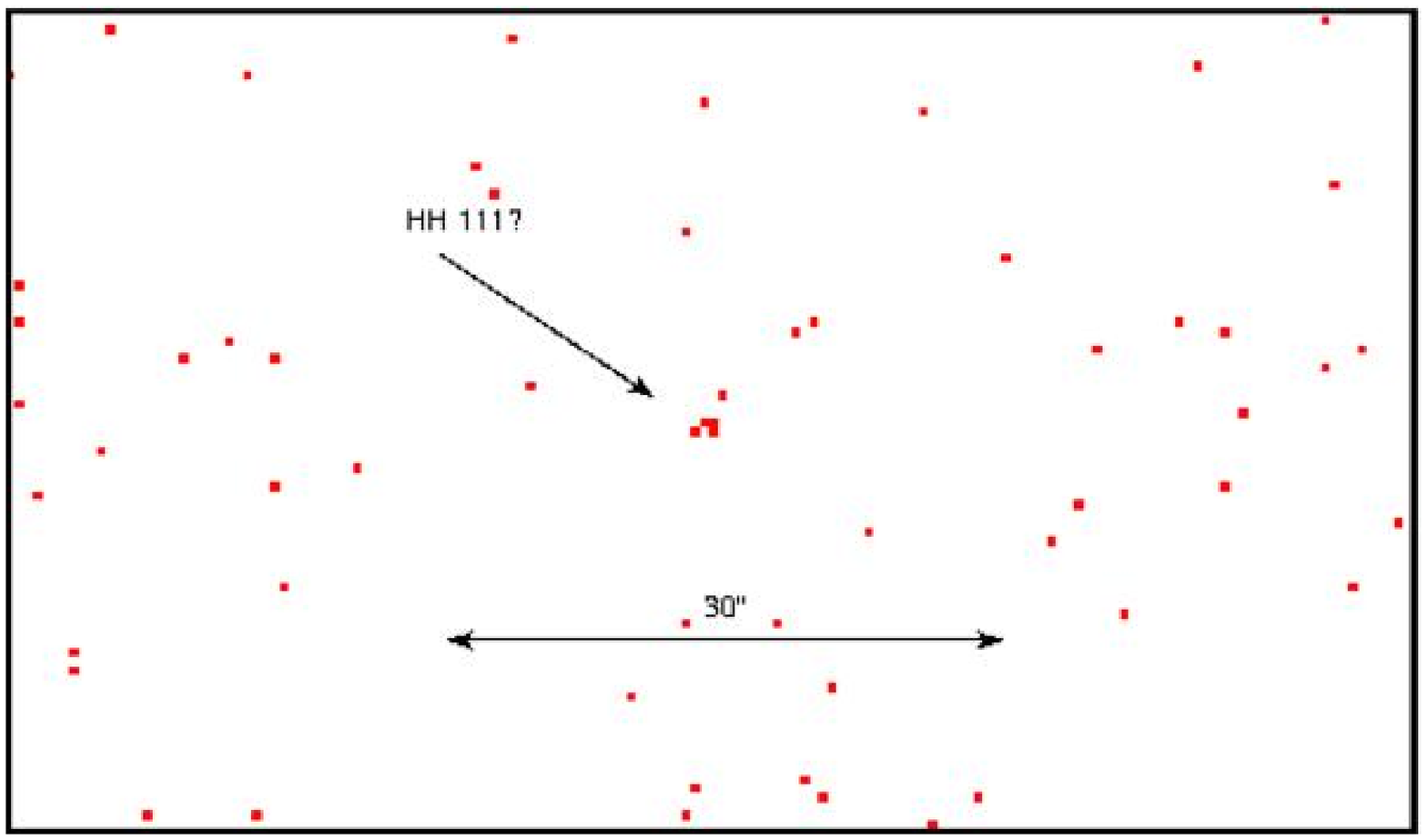}}
\caption{{\it Chandra} ACIS-S images of jet-driving sources. Pixel size is $0.49\arcsec$; all images were exposed for
$\approx 30$~ks and show counts in the 0.5--7~keV range except for HH~111, where only counts in the 3--7~keV
range are shown to minimize background. From top to bottom:   
         DG Tau A and B (separation $\approx 50^{\prime\prime}$); 
	 FS Tau A and B (separation $\approx 20^{\prime\prime}$); 
	 HH34 IRS; HH111 IRS (tentative detection).
	 \label{images}}
\end{figure}

Three of the four target sources (DG Tau B, FS Tau B, and HH~34 IRS) were clearly detected very close to 
the expected positions  based on {\sl wavdetect}, while only 4 counts 
were registered for HH~111 (Fig.~\ref{images}).  The close positional coincidence with the radio source 
(0.15\arcsec\ distance from the HH~111 radio position given by \citealt{rodriguez94}), the strong clustering of 
the counts within a radius of 1--2 pixels (Fig.~\ref{images}), the energy of the four counts (3.3--6.4~keV, very
similar to the distributions found for the other absorbed sources), and the rather low background 
level make it probable that these photons originate from the HH~111 jet-driving
 protostar although the significance is too low to  prove this identification. The background level 
 was such that 0.1 counts would be expected in the 3--7~keV range in a circle with a radius of 3\arcsec\ 
around the source, and 0.6~cts in the 0.3--10~keV range.

\begin{table*}[t!]
\centering
\caption{Protostellar, jet-driving X-ray sources}
\begin{tabular}{lrrrrrrrr}
\hline
\noalign{\smallskip}
{\rm Parameter}                           &  DG Tau B	& Ref$^a$   & FS Tau B   &  Ref$^a$   & HH 34 IRS	&  Ref$^a$  & HH 111 IRS  &  Ref$^a$  \\
\noalign{\smallskip}
\hline
\noalign{\smallskip}
Useful spectral range (keV)               & 2--6.5      & -	    & 2--7	 & -	      & 3--7	    & -	    & 3--6.5      & -	   \\
Detected X-ray counts                     & 9		& -	    & 32	 & -	      & 15	    & -	    & 4	          & -	   \\
Position:                                 &		&	    &		 &	      & 	    & 	    &	          &	   \\
\ Expected R.A. (h~m~s)                   & 04 27 02.55	& 1	    & 04 22 00.70& 2	      & 05 35 29.84 & 2	    & 05 51 46.31 & 3	   \\
\ Expected $\delta~(\deg\ \arcmin\ \arcsec)$& 26 05 30.90& 1        & 26 57 32.50& 2	      &-06 26 58.40 & 2	    & 02 48 29.72 & 3	   \\
\ Observed R.A. (h~m~s)                   & 04 27 02.58	& -	    & 04 22 00.71& -	      & 05 35 29.84 & -	    & 05 51 46.30 & -	   \\
\ Observed $\delta~(\deg\ \arcmin\ \arcsec)$& 26 05 30.91& -	    & 26 57 32.17& -	      &-06 26 58.24 & -	    & 02 48 29.70 & -	   \\
\ Offset (\arcsec)                        & 0.40	&	    & 0.36	 & -	      & 0.16	    & -	    & 0.15        & -	   \\
X-rays:                                   &       	&	    &    	 & -	      &    	    &  	    &             &	   \\
\ Expected $N_{\rm H}^b$                  & 1.4      	& 6   & $>1.6$, $\ga 2$, 4.6& 1, 6, 4 & 1   	    & 5     &  $>6$       & 7	   \\
\ Measured $N_{\rm H}^b$                  & 51 (6--72)  &	    & 26 (10--37) & -	      & 28 (14--54) & -	    & -           & -     \\
\noalign{\smallskip}
\hline
\multicolumn{9}{l}{$^a$ References: 1 = \citet{padgett99}, 2 = 2MASS, \citet{cutri03}, 3 = \citet{rodriguez94}, 4 = \citet{krist98},}\\ 
\multicolumn{9}{l}{5 = \citet{reipurth86}, 6 = \citet{mundt87}, 7 = \citet{reipurth89}}\\
\multicolumn{9}{l}{$^b$ Expected $N_{\rm H}$ from reported optical extinction: $N_{\rm H} \approx 2\times 10^{21}A_{\rm V}$~cm$^{-2}$}\\
\multicolumn{9}{l}{$N_{\rm H}$ is given in units of $10^{22}$~cm$^{-2}$, 90\% confidence ranges are in parentheses}\\
\end{tabular}
\label{proto}
\end{table*}

\begin{figure*}
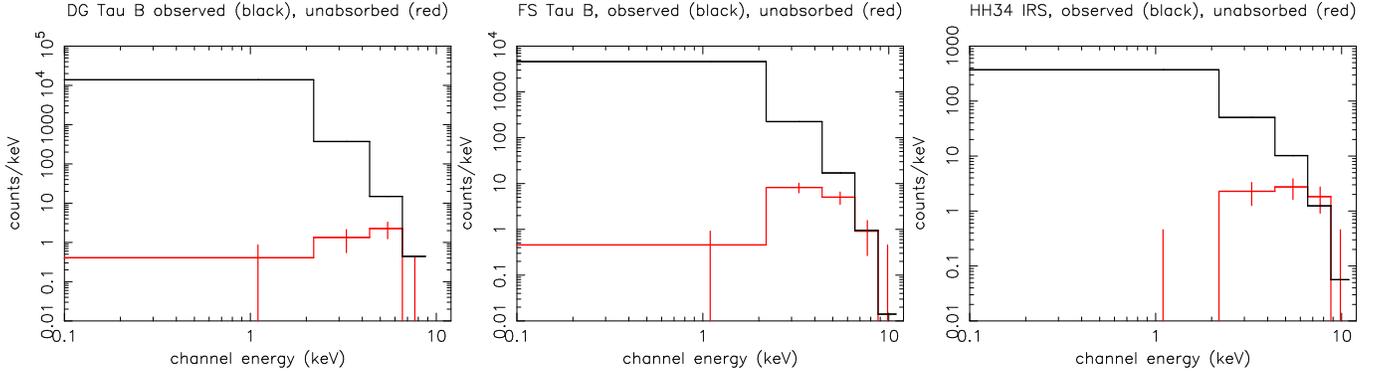

\hbox{
\resizebox{0.33\hsize}{!}{\rotatebox{270}{\includegraphics{guedel_f6a.ps}}}
\resizebox{0.33\hsize}{!}{\rotatebox{270}{\includegraphics{guedel_f6b.ps}}}
\resizebox{0.33\hsize}{!}{\rotatebox{270}{\includegraphics{guedel_f6c.ps}}}
}
\caption{Binned ACIS-S CCD spectra (red/gray, with error bars) of (from left to right) DG Tau B, FS Tau B, and HH34 IRS, 
compared with the modeled, response-folded intrinsic spectra if the photoelectric absorption is removed (black solid histograms). 
\label{spectra}}
\end{figure*}

To avoid systematic effects in the spectral fits due to coarse binning of the few counts, we
used the C statistic in XSPEC in combination with unbinned data in the energy range
of 0.5--8~keV.  
The  metallicity was fixed at  0.2 times solar photospheric values 
(\citealt{anders89}), coincident with the adopted Fe abundance for the other spectral fits (we
note that the only emission lines significantly contributing to the absorbed protostellar
sources are those in the Fe complex at 6.7~keV; even adopting solar abundances did not change 
our results significantly). 
For illustration purposes, we show in Fig.~\ref{spectra} binned spectra together with
the best-fit {\it unabsorbed} models. These figures illustrate that most of the flux
is suppressed by photoelectric absorption.  
The best-fit results and the acceptable 90\% ranges for $N_{\rm H}$ are summarized in Table~\ref{proto}; other
parameters  were not sufficiently constrained for further discussion.

\section{Discussion}\label{sect:discussion}

The TAX spectra from the optically revealed jet-driving T Tau stars 
presented in this paper clearly require two spatially distinct sources because both their
time behavior and their absorption along the line of sight are different. We now discuss 
possible models.

\subsection{CTTS: The hard component}

\begin{figure}[t!]
   \centering
   \includegraphics[angle=-0,width=8.cm]{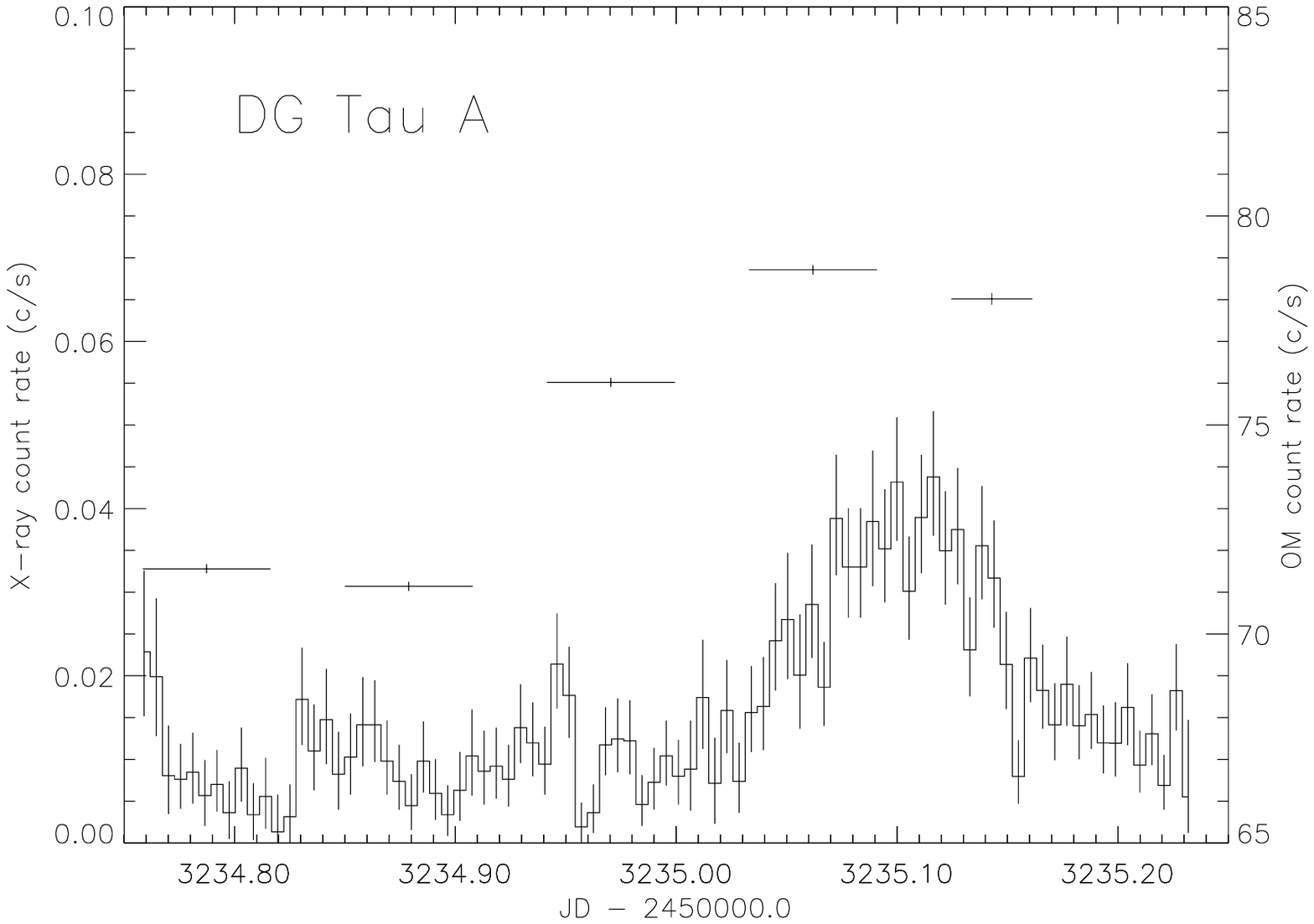}
   \vskip 0.3truecm
   \includegraphics[angle=-0,width=8.cm]{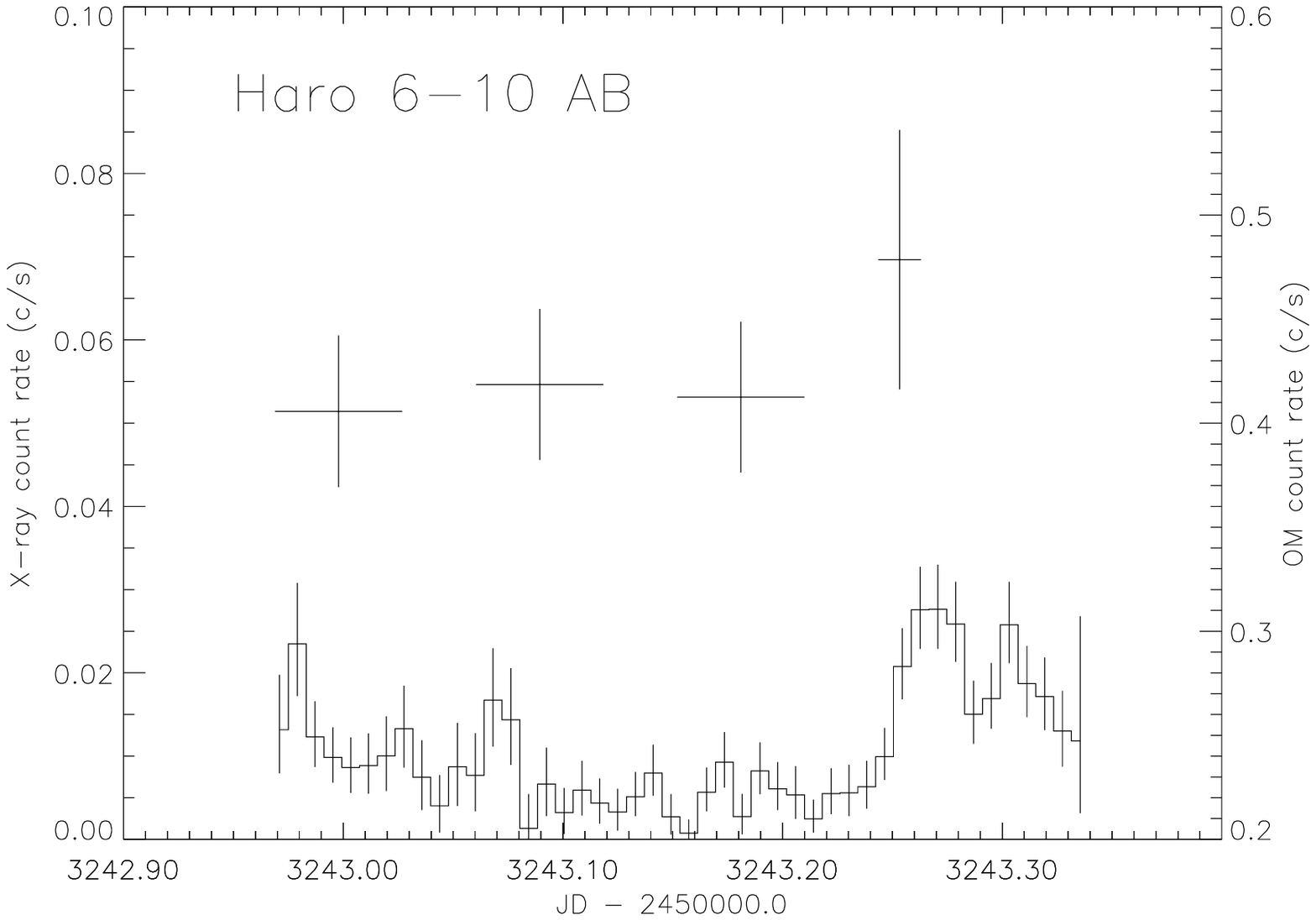}
   \caption{Correlation between X-ray emission and U band emission. The 
   U band data are shown by the large crosses above the X-ray light curves and 
   refer to the y axis labels on the right side.
   {\it Top:} DG Tau A; {\it bottom:} GV Tau.}
              \label{omx}%
    \end{figure}

The hard components of the spectra reveal very high temperatures. Such temperatures are commonly 
found in other T Tau stars \citep{skinner03, telleschi06}. The  flares seen in the light curves 
clearly argue in favor of a magnetic origin. For DG Tau A,
the U band light curve extracted from the simultaneous observations with the OM
peaked before the X-rays, in  analogy to 
solar flares (Fig.~\ref{omx}, upper panel). We therefore interpret this component as being 
due to a magnetically confined corona above the surface of the star.
The time resolution of the OM data is too coarse to reveal a clear temporal relation for GV Tau
(Fig.~\ref{omx}, lower panel).

The excessive absorption of this coronal component requires a large amount of cool gas along the line of
sight to the observer. A cool wind or a molecular outflow are   potential absorbers. However, in that case the soft component
must be located at very large distances from the star in order to escape from strong absorption. Also, dust admixtures
would lead to correspondingly strong optical extinction of the star, which is not observed.

We propose  a scenario related to accretion streams. The mass accretion rates of our
stars are among the largest found in any TTS in the TMC 
(Table~\ref{params}). Much of the gas accreting from the circumstellar disk is only
weakly ionized. This gas will stream along the 
magnetic field lines that form the corona and thus enshroud X-ray bright coronal 
loops so that photoelectric absorption attenuates the soft X-rays from the underlying 
coronal plasma.

Why, then,  is the optical extinction
of the stars relatively small? The measured $N_{\rm H}$ columns  
would imply $A_{\rm V} \approx 10-20$~mag for each star while measured optical
extinctions are only a few magnitudes (Table~\ref{params}). Such deviations are possible 
if the X-rays are propagating through an anomalous mixture of gas and dust in which dust is
depleted. Dust destruction and evaporation occurs at temperatures around 
$T_{\rm sub} \approx 1600-2000$~K \citep{dalessio98, whitney04}, and the dust sublimation radius 
from the center of the star can be estimated from the empirical formula,
\begin{equation}
{R_{\rm sub}\over R_*} = \left({T_{\rm sub}\over T_{\rm eff}}\right)^{-2.085}
\end{equation}
(\citealt{whitney04}, where $T_{\rm sub} = 1600$~K has been assumed). Using $T_{\rm eff}$ 
from Table~\ref{params}, we find radial distances 
of $7-10~R_*$. As a consequence, there should be no strong extinction by dust from the accretion 
streams while photoelectric absorption remains in effect. In this picture, visual and 
near-infrared extinctions are due to the larger-scale environment of the star, while 
photoelectric absorption is due to both the large-scale environment (for the 
soft components) and, more strongly, the immediate surroundings of the star 
(for the hard component). Our observations may thus give indirect evidence for 
dust sublimation in the inner circumstellar accretion disks of T Tau stars on the one
hand and for strong gaseous accretion streams in the innermost region or onto the star on the other hand.

\subsection{CTTS: The soft component}

The soft component dominates the spectrum below 1~keV. The spectral shape in this
region shows no indication of an equally prominent plasma component at temperatures 
$\ga 10$~MK. This component, considered in isolation, is unusual for T Tau stars
that normally reveal spectra dominated by hot temperatures \citep{telleschi06}.
The  exception  extensively discussed in the recent literature is
TW Hya, the X-ray spectrum of which is strongly dominated by a soft spectral component of this kind. TW Hya is,
however, a rather evolved classical T Tau star with an age of $\approx 10$~Myr and
a low accretion rate of $4\times 10^{-10}~M_{\odot}$~yr$^{-1}$ \citep{muzerolle00}. 
\citet{kastner02} proposed that
the low temperatures and the high electron densities inferred from the O~VII He-like
triplet are indicative of X-rays produced in accretion shocks rather than coronal
X-rays. \citet{stelzer04} argued that the anomalous abundances found in the TW Hya
X-ray spectrum are further support for this picture.

In the case of the jet-driving sources, this explanation faces serious difficulties.
The accretion rates measured for these stars are among the highest known in TMC.
If X-rays are produced near the footpoints of accretion funnels, then we would expect
that the same gas streams absorbing the hot coronal emission also strongly absorb
the soft component.

On the contrary, the measured absorption column densities tend to be smaller than
expected from the optical extinction  of the star, at least for DG Tau A. Either, the soft X-rays escape 
through a region of depleted gas in between the accretion streams, or they are 
formed significantly outside the immediate environment of the star. Consulting 
Table~\ref{params},  the distinguishing property of the TAX sources are 
 their jets. The strong differences in $L_{\rm X}$ of the cool component
(ratio of 1:0.56:0.04 for DG Tau A : GV Tau : DP Tau)  mirror in two
parameters: the equivalent width of [O~I] (1:0.23:0.05), and the 
mass outflow rate (1:0.3:0.06). Although no linear relation is to be expected,
the trend is rather suggestive for outflow-related X-ray activity.

 A faint soft component was also found at a distance of 2--5\arcsec\ from DG Tau A, 
co-spatial with  bow-shock structures in the jet (\citealt{guedel05}, see also Fig.~\ref{images}). 
Because the knots at this distance show X-ray emission 
while those further away do not, \citet{guedel05}  proposed that the more luminous
soft component forms in shocks even closer to the star, presumably at the base of the jets but
clearly outside the magnetospheric accretion zone around the star. This would
simply require that the jet portions closer to the star contain more emission measure heated 
to X-ray temperatures although the temperatures would be the same. This scenario  is supported by
 the close similarity between the soft component of the X-ray spectrum obtained 
from the region within one arcsecond around the star, and the extended X-ray source coincident with
the jet.

Further support for this hypothesis comes from optical extinction measurements of the
Herbig-Haro structures in the jet of DG Tau A. \citet{cohen85} found an $A_{\rm V}$ of
0.39~mag while the stellar $A_{\rm V}$ has been reported between 1.4~mag and 3.3~mag (Table~\ref{params}).
Using standard interstellar gas-to-dust ratios, we would expect a hydrogen column density of
$N_{\rm H} \approx 8 \times 10^{20}$~cm$^{-2}$ toward the jet, comparable with the X-ray determined value of 
$(1.1\pm 0.3) \times  10^{21}$~cm$^{-2}$. 

The precise cause for the 
heating of such jet/related X-ray sources is unclear. \citet{bally03} suggested, for an X-ray 
source in the jet but spatially resolved from the location of the fully absorbed 
protostar L1551 IRS-5, a number of scenarios:

i) Shock-heating at the working surface of a bow shock colliding with the
ambient medium, in analogy to shock heating in the more distant HH objects; 
ii) Thompson scattering of protostellar X-rays  into the line of sight by a cool, 
dense medium  perhaps in the outer, expanding regions of the circumstellar disk 
or  the accreting envelope;
iii) shock heating at an obstacle immersed in the jet flow, where either magnetic fields 
or a dense, ambient medium  redirects  the initially expanding wind and collimates it 
to a jet;
iv) wind-disk or wind-wind collisions in a binary system.
 We are not in a position to discriminate
between such models, except to note that the final option  is unlikely to
apply to our sources, as discussed in the  next section.

We speculate that magnetic fields may further heat the jet gas. The jet of DG Tau A
has been found to rotate around its axis. This motion derives from the Kepler motion of
the circumstellar disk from where the jet accelerates. The motion is differential in
radius, and the mass flow along the field generates a trailing spiral \citep{anderson03}. In ionized gas, 
such field configurations induce currents  that may dissipate and heat the gas if its resistivity is large enough.
Further irregularities can be introduced by pulsed outflows, precession, and the shock with the ambient
medium \citep{cerqueira04}. Magnetic ``tangential discontinuities'' may lead to reconnection, thus
liberating additional (magnetic) energy \citep{parker83}. Direct magnetic field measurements are 
yet to be made, although
radio observations suggest the presence of weak magnetic fields in jets or outflows (e.g., \citealt{ray97}). 

\begin{figure}[t!]
   \centering
   \includegraphics[angle=-90,width=8.6cm]{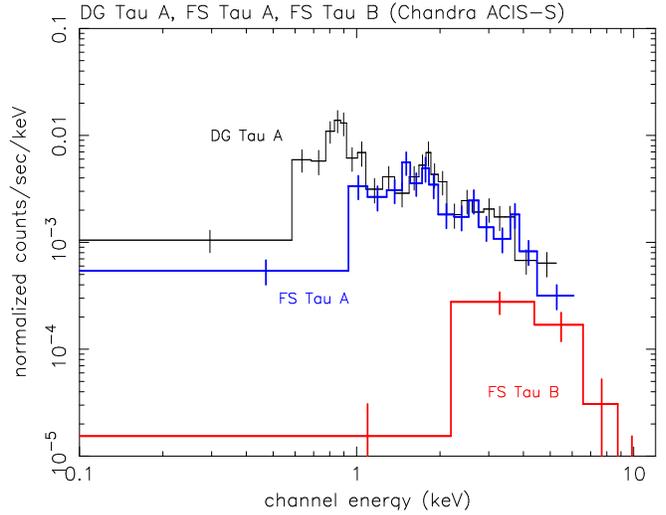}
   \caption{Comparison of the three {\it Chandra} spectra from DG Tau A (black, with soft component),
             FS Tau A (blue, thick), and FS Tau B (red).}
              \label{comparison}%
    \end{figure}

We next compare in Fig.~\ref{comparison} three CCD spectra from {\it Chandra} (ACIS-S): 
The spectrum of DG Tau A (black, with soft excess), the spectrum of FS Tau A (thick, blue),
and the spectrum of the protostar FS Tau B.\footnote{we discuss data only from {\it Chandra} for this 
comparison  because the strongly absorbed  FS Tau B is located in the wings of the
{\it XMM-Newton} image of FS Tau A, and FS Tau A itself was showing a gradual decay from
a strong flare during the {\it XMM-Newton} observation} FS Tau A is subject to 
a similar absorption  to the hot component in DG Tau A ($N_{\rm H} = [0.9-1.4]\times 10^{22}$~cm$^{-2}$, 
\citealt{guedel06}), shows a similar (unabsorbed) $L_{\rm X} \approx 3.4\times 10^{29}$~erg~s$^{-1}$, and
also reveals an extremely hot electron temperature of 43~MK (higher temperature of  a 
two-component  fit; \citealt{guedel06}). FS Tau is, however, not known to drive strong jets, although
a poorly collimated wind seems to be present \citep{hirth97}.
While the two spectra are very similar above 1.5~keV, the absence of a soft component
in FS Tau A is striking. The spectrum of the jet-driving FS Tau B does not show any soft emission 
because such emission from the star and the jet  within an arcsecond or so would be absorbed.
The hard component above 5~keV, however, is again similar in flux to the hard spectral component
of the other two stars, suggesting a similar coronal component.

 Optical jets are detected in predominantly low-excitation lines such as forbidden lines
of [O~I], [N~II], and [Si~II] formed at no more than a few thousand degrees \citep{hirth97}.
On the other hand, the X-ray emission detected in direct imaging around DG Tau A \citep{guedel05}
and the soft spectral components suggested here to be due to jets as well require plasma with
electron temperatures above 1~MK. The challenge is to
reconcile the observational evidence from two widely disparate temperature ranges
when the presence of an intermediate temperature range ($10^4 - 10^6$~K) is apparently less 
evident in the observations.  What
is the status with respect to intermediate-excitation lines?

First, detectability of forbidden lines requires favorable
densities for their formation, and a sufficient emission measure.
Modeling of [O~I], [N~II], and [Si~II]  lines indicates
that they form in distinct regions \citep{hamann94}; [O~I] forms in a gas in 
the electron-temperature range of 
9,000--14,000~K and high densities ($n_{\rm e} \approx 5\times 10^5 - 10^7$~cm$^{-3}$),
while [Si~II] forms above 13,000~K and in lower densities ($n_{\rm e} 
\approx 10^3 - 7\times 10^4$~cm$^{-3}$). At high velocities only, [N~II] and
[O~II] lines may be formed, requiring $T_{\rm e} \ga 15,000$~K and
$n_{\rm e} \la 10^5$~cm$^{-3}$. These results therefore indicate
that $n_{\rm e}$ decreases away from the star, while $T_{\rm e}$  increases,
at least as far as high-velocity gas is concerned.

Conditions in the jets may in general simply not be favorable to the production of 
detectable amounts of flux in higher-excitation lines, e.g., [O~III]. On the 
other hand, there are now several reports  of the presence of hot winds and 
outflows from T Tau stars. \citet{beristain01} presented an extensive study of
He~I and He~II emission line profiles. The important diagnostic power of these
lines lies in their high excitation potentials (20--50~eV), 
requiring gas temperatures of $(2.5-9)\times 10^4$~K in the case
of collisional excitation \citep{beristain01}. The authors found 
that blueshifted absorption features in broad He~I line profiles
and maximum blue-wing velocities exceeding 200~km~s$^{-1}$ indicate
the presence of {\it hot, coronal winds} in about half of their sample, 
one of the clearest examples being DG Tau A with a maximum blue-wing
velocity of 600~km~s$^{-1}$. These winds would be launched in the polar
region of the stars, complementary to cooler winds  that are probably 
accelerated around the disk/magnetosphere boundary.

Work on He lines was extended to the He~I $\lambda$10830 line by \citet{takami02},
\citet{edwards03} and \citet{edwards06}. This feature is again a high-excitation 
line (20~eV, requiring $T \approx 10^4$~K for collisional ionization) 
and has an unprecedented sensitivity to hot, inner winds. Significant
absorption below the continuum was found  in 71\% of the CTTS, but in 
none of the WTTS, suggesting that the feature is due to accretion-driven
winds, originating in the immediate vicinity of the star. Specifically in the
case of DG Tau A, absorption and emission components were found, identified 
with an inner wind and the jet shocks, respectively \citep{takami02}. 
Further, [O~III]~$\lambda$4959 and $\lambda$5007 emission is indeed also found 
in one of the outer working surfaces of the DG Tau jet \citep{cohen85} and in
other shock-related features in Herbig-Haro objects (\citealt{matt03} and references 
therein),  indicating that singly ionized O is further ionized beyond its ionization 
potential of 35~eV. Shock modeling suggests post-shock electron temperatures in 
the range of $(6-7)\times 10^4$~K \citep{cohen85}, and higher temperatures at
specific locations in Herbig-Haro shock regions must exist to explain recently
detected X-ray emission (see summary in Sect.~\ref{introduction}).
 
\citet{gomez01} have analyzed high-resolution profiles of semiforbidden ultraviolet
lines of C~III]  $\lambda$1908 and Si~III] $\lambda$1892 in two CTTS. They concluded that
the lines are not formed in accretion shocks but in a hot wind outside $2R_*$ and
not farther from the star than 38$R_*$ (for RY Tau), suggesting a bow-shaped
shock source at the base of the jet, with densities $>10^5$~cm$^{-3}$ (i.e., at least
$10^2-10^3$ times denser than Herbig-Haro objects) and temperatures in 
the range of $5\times 10^4 - 10^5$~K.  

Still higher outflow temperatures  have been suggested  by  far-ultraviolet 
observations. The CTTSs TW Hya and T Tau both show P Cygni profiles, line
asymmetries, and absorption in C~III $\lambda$977 and O~VI $\lambda$1032 lines
that are indicative of a fast (400~km~s$^{-1}$) and hot ($3\times 10^5$~K) accelerating
outflow close to the star \citep{dupree05}.

In summary, although there is no coherent model of winds and jets at all
temperature and density levels, it appears clear that CTTS show evidence for
jet- or wind-related gas at temperatures from $< 10^4$~K to several times $10^5$~K.
It is then natural to think of the X-ray emission as a continuation toward higher
temperatures. It appears that at various locations in the jet, in particular 
in shock regions, conditions may be favorable for the production of X-rays, while
densities and volumes seem to be unfavorable for appreciable line emission at
lower ionization stages such as [O~III]. A speculative possibility would be
that rapidly heated X-ray sources in the jets freely expand and cool in such a way
that their densities and emission measures are too low for easy detection
in intermediate-excitation lines once the temperature  has dropped to appropriate
values. The dominant [O~I] emission is not necessarily formed at the same locations 
\citep{hamann94}.

\subsection{Binarity?}

\begin{figure*}[t!]
   \hbox{
   \includegraphics[angle=-0,width=5.7cm]{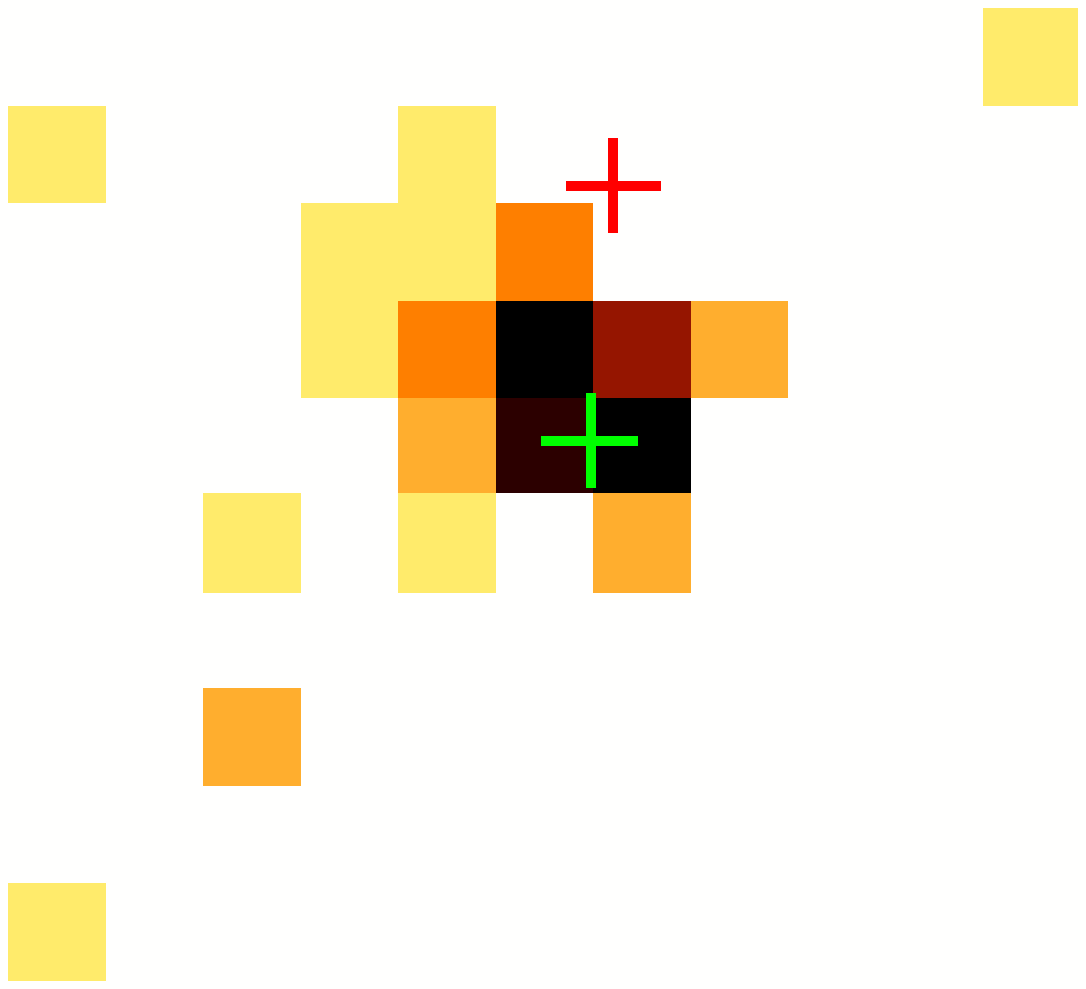}
   \includegraphics[angle=-0,width=5.7cm]{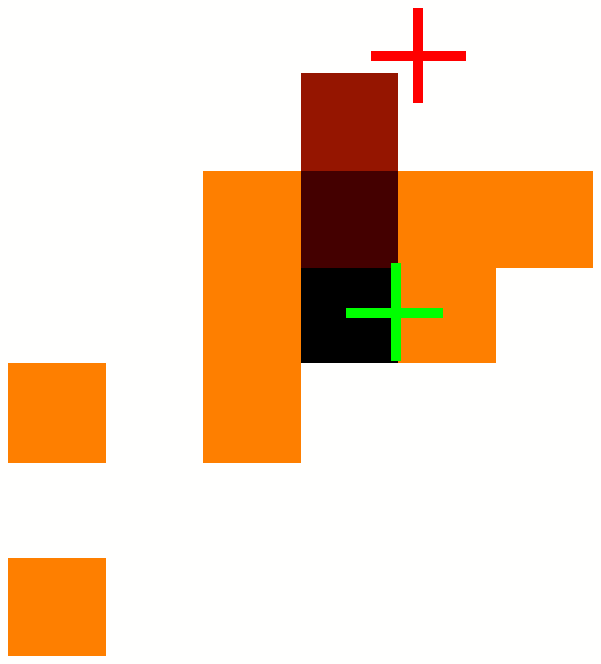}
   \includegraphics[angle=-0,width=5.7cm]{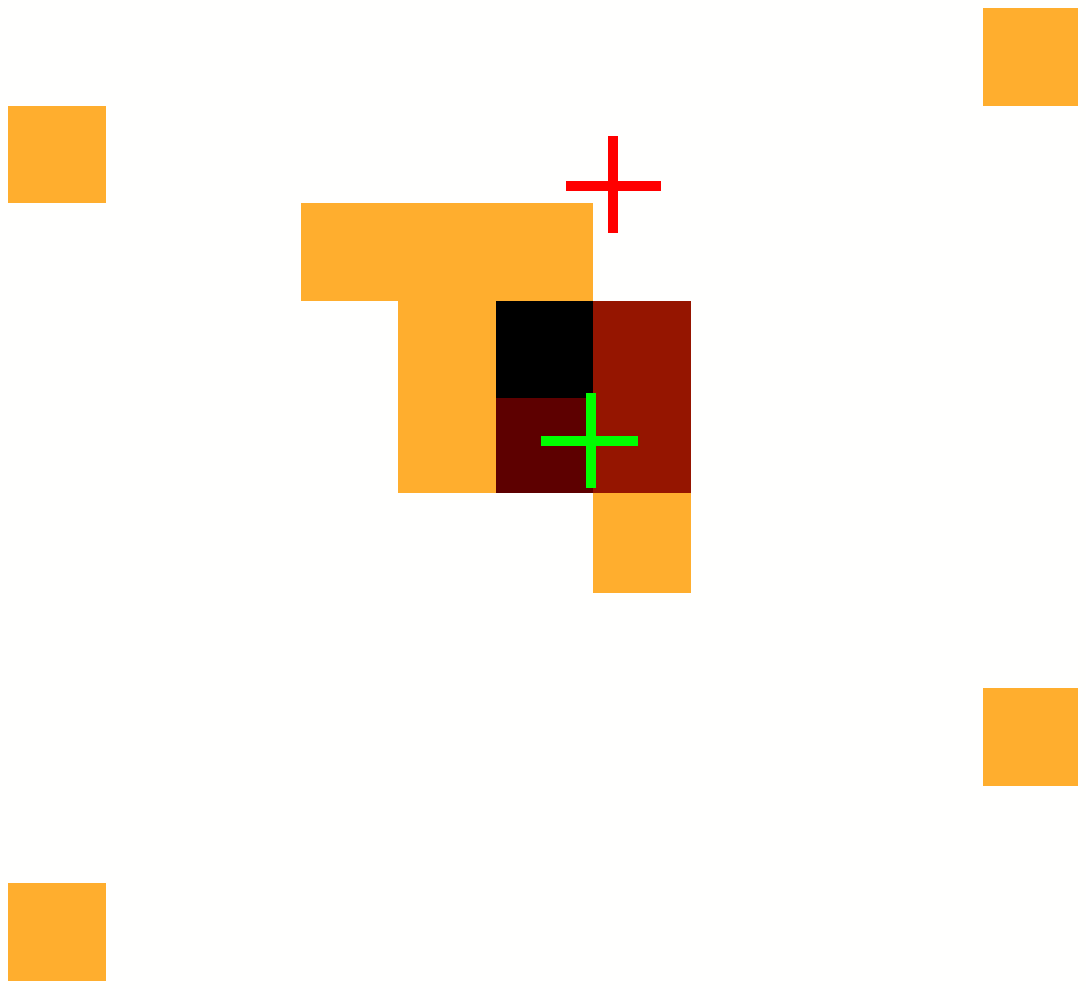}
   }
   \caption{{\it Chandra} images of GV Tau. {\it Left:} Counts in the range 0.5--5~keV.
   The lower cross marks the 2MASS position of GV Tau A, the upper cross  the
   GV Tau B protostar, applying the same offset to GV Tau A as measured in radio
   data \citep{reipurth04}. {\it Middle:} Similar, for the 0.5--1~keV range. {\it Right:} Similar,
   for the 2.2--7~keV range. Pixel size is $0\farcs 492$.}
	      \label{GVTauChandraIm}%
    \end{figure*}

TAX spectra may originate from a binary with components that 
are located behind largely different gas columns, each thus contributing one of the spectral 
components. This model is unlikely for the following reasons: i) The less absorbed companion 
would in all cases reveal a uniquely soft, non-flaring X-ray component only. None of the bright
TTS in the XEST survey, single or multiple, revealed such an X-ray spectrum \citep{telleschi06, guedel06}.
ii) Except for GV Tau,  no companions have been found to the TAX sources despite detailed
searches  (e.g., \citealt{leinert91}). iii) GV Tau = Haro 6-10 is indeed a binary, the more 
absorbed component being an embedded protostar behind a large gas column  (see 
\citealt{reipurth04} for a radio image). A high-resolution
archival {\it Chandra} image (ObsID = 4498), however, reveals that the soft and hard photons 
originate from the same location, and  this location agrees, within the error ranges
for {\it Chandra}, the VLA \citep{reipurth04} and 2MASS, with the less absorbed component 
Haro 6-10A (Table~\ref{GVcoords}; Fig.~\ref{GVTauChandraIm}). We conclude that the embedded Haro 6-10B does not 
contribute to the detected X-ray emission.

We also  note that the prominent jet from the GV Tau system is  associated
with the optically revealed star GV Tau A rather than the deeply embedded infrared source
\citep{reipurth04}.

\begin{table}[t!]
\centering
\caption{Coordinates of GV Tau A, B}
\label{GVcoords}
\begin{tabular}{llll}
\hline
\noalign{\smallskip}
            Object      &   Observation           & RA(J2000.0)$^a$ & $\delta${\rm (J2000.0)}$^a$ \\
                        &                         & h\ m\ s      &     $\deg\ \arcmin\ \arcsec$  \\
            \noalign{\smallskip}
            \hline
            \noalign{\smallskip}
            GV Tau A     &    2MASS     &  4  29  23.734    &  +24  33  00.277 \\
            GV Tau B$^b$ &    2MASS     &  4  29  23.725    &  +24  33  01.577  \\
            \noalign{\smallskip}
            \hline
            \noalign{\smallskip}
            GV Tau A$^c$ &   VLA       &  4 29 23.733    &  +24 33 00.125 \\ 
            GV Tau B$^c$ &   VLA       &  4 29 23.724    &  +24 33 01.425 \\ 
            \noalign{\smallskip}
            \hline
            \noalign{\smallskip}
            soft X  & {\it Chandra}        &  4 29 23.744(8) &  +24 33 00.54(11)  \\
            hard X  & {\it Chandra}        &  4 29 23.749(7) &  +24 33 00.66(10)  \\
            total X & {\it Chandra}        &  4 29 23.737(4) &  +24 33 00.53(5)      \\
            
            \hline
         \end{tabular}
\begin{list}{}{}
\item[$^{\mathrm{a}}$] Errors for {\it Chandra} sources in units of last digits; formal {\sl wavdetect} fit errors
\item[$^{\mathrm{b}}$] Applying radio offset to GV Tau A 2MASS source
\item[$^{\mathrm{c}}$] From \citet{reipurth04}
\end{list}
\end{table}

\subsection{The protostars}

 Although the ACIS protostellar sources are very faint, their photon energy distribution 
is revealing: we detect at most one count below $\approx 2$~keV in each source, which is 
normally the dominant spectral range for a coronal source.

The  emission cannot be radiation from shocks in the jets or 
Herbig-Haro objects. Such sources are too soft and will therefore be absorbed 
(\citealt{bally03}). The hard sources required here are reminiscent of coronal emission 
of very active stars.  Useful information on the hydrogen absorption column density, 
$N_{\rm H}$, can therefore be obtained {\it for the circumstellar disk}.

Although the 90\% confidence ranges in $N_{\rm H}$, $kT$,  and $L_{\rm X}$ are large
(Table~\ref{proto}), the values of $N_{\rm H}$ 
are  consistently found in the range of $\approx (5-50) \times 10^{22}$~cm$^{-2}$ if we require ``reasonable'' 
values for  $kT$ and $L_{\rm X}$ (e.g., typical saturation values given the estimated 
$L_*$, and $T$ in the range of $6-60$~MK). We thus expect $A_{\rm V} \approx 30-300$~mag, 
much higher than reports in the literature ($A_{\rm V} \approx 5-25$; 
\citealt{reipurth86}, \citealt{reipurth89}, \citealt{mundt87}, \citealt{krist98}, P99):
Either, the $A_{\rm V}$ values are erroneous due to IR emission from the stellar surrounding, or   
the gas-to-dust (G/D) ratios are significantly enhanced  in these circumstellar disks.

\section{Summary and conclusions}\label{sect:summary}

Jets are assumed to be launched within a few AU from the central star and have
been resolved down to 0.1\arcsec\ (14~AU) from the latter at the distance of the Taurus
Molecular Cloud  \citep{bacciotti00}. These regions remain unresolved
from the central star in present-day X-ray observations. Nevertheless, several
jet-driving T Tau stars reveal peculiar X-ray spectra that have been taken as 
evidence for shock formation at the base of the jets (\citealt{guedel05}, and our
discussion above).
In all cases, a very weakly attenuated ($N_{\rm H} \approx 
10^{21}$~cm$^{-2}$), luminous but very soft spectral component indicates electron 
temperatures of  no more than a few MK. While this component produces photons mostly 
between 0.5--1~keV, a strongly absorbed ($N_{\rm H} > 10^{22}$~cm$^{-2}$), hard spectral 
component dominates above 2~keV, resulting in a shallow spectral shape with two peaks.
The hard component  originates from hot plasma of several tens of MK. In each case, it 
reveals flares while no significant  variability is detected from the soft component  whose flux
is approximately  constant, even on time scales of 6--8 months (Sect.~\ref{longterm}). 

We attribute the hard component to a magnetospheric corona, as observed on most T Tauri stars. Although
these T Tauri stars are unusually strong accretors (with $\dot{M}_{\rm acc}$ up to 
$\approx 10^{-6}~M_{\odot}$~yr$^{-1}$) we do not attribute the soft component 
to an accretion shock close to the surface of the star as proposed for the so far unique CTTS TW Hya
\citep{kastner02}. Such emission would be subject to the same high absorption experienced
by the hard component, while the observed soft component is little absorbed. Instead, we propose
that the soft component originates further away from the star where the first shocks form at the base of the jets.
Indeed, the X-ray luminosity of the soft component in these three stars appears to scale with the mass 
outflow rate and with the equivalent width of [O~I] from the jet.

It is interesting to note that the visual extinction of the T Tau stars showing Two-Absorber X-ray (TAX) spectra
is modest and does not agree with the photoelectric absorption of the hard component if standard gas-to-dust
mass ratios compatible with the interstellar medium are assumed. 
We suggest that the hard emission is absorbed by accreting material an that this accreting material
is dust-depleted. This is supported by the dust
sublimation radius of our sources being several stellar radii. Our observations thus
provide both evidence for dust sublimation in the innermost part of the accretion disk and
for massive gas accretion streams from the disk to the star enshrouding the hot coronal 
plasma. Because the line of sight toward DG Tau A is inclined against the axis at a rather 
small angle of $37.7\pm 3$~deg \citep{eisloeffel98}, the absorbing mass accretion streams 
must reach high stellar latitudes rather than fall toward the equatorial regions. This fully 
supports the common picture of magnetic accretion (e.g., \citealt{calvet98}).

Detection of the soft jet component requires low attenuation
by gas, i.e., it is inaccessible in embedded protostars although such X-rays may
locally exist in all jet-driving sources. We have indeed found only hard components
in our strongly obscured targets, and they coincide very closely with the position of
the stars. No displaced soft components were identified although we recall that a soft jet source
was reported previously for DG Tau A  (\citealt{guedel05}; see also Fig.~\ref{images}).

\begin{acknowledgements}
We thank several of our colleagues, in particular  Nicolas Grosso, Fran\c{c}ois M\'enard, and 
Karl Stapelfeldt, for discussions on the subject of this paper. We acknowledge constructive
comments by an anonymous referee that helped to significantly improve this paper. 
We warmly acknowledge financial support by the International Space Science Institute (ISSI) in Bern to
the {\it XMM-Newton} XEST team. X-ray astronomy research at PSI has been supported by the Swiss National Science
Foundation (grants 20-66875.01 and 20-109255/1). MA acknowledges support by NASA grants NNG05GF92G.
Part of this research is based on observations obtained with {\it XMM-Newton}, an ESA science mission
with instruments and contributions directly funded by ESA member states and the USA (NASA).
This publication makes use of data products from the
Two Micron All Sky Survey (2MASS), which is a joint project of the University of Massachusetts
and the Infrared Processing and Analysis Center/California Institute of Technology,
funded by the National Aeronautics and Space Administration and the National Science  
Foundation. Further, our research has made use of the SIMBAD database,
operated at CDS, Strasbourg, France. 
\end{acknowledgements}

\end{document}